\documentclass[conference]{IEEEtran}
\IEEEoverridecommandlockouts
\usepackage{common}

\newenvironment{acks}
{\section*{Acknowledgments}
}
{}

\newcommand{\city}[1]{#1\xspace}
\newcommand{\country}[1]{, #1\xspace}

\def\BibTeX{{\rm B\kern-.05em{\sc i\kern-.025em b}\kern-.08em
    T\kern-.1667em\lower.7ex\hbox{E}\kern-.125emX}}
\begin{document}

\title{\ourtitle}

\ifanonymous
  \iftrue
    \author{\IEEEauthorblockN{\nopx}
    \IEEEauthorblockA{\textit{\ksrl} \\
    \textit{\kit}\\
    \ka}
    \and
    \IEEEauthorblockN{\lukas}
    \IEEEauthorblockA{\textit{\ksrl} \\
    \textit{\kit}\\
    \ka}
    \and
    \IEEEauthorblockN{\chris}
    \IEEEauthorblockA{\textit{\ksrl} \\
    \textit{\kit}\\
    \ka}
    }
  \else
    \author{\begin{minipage}[t][20mm][c]{\textwidth}
    \centering
    --- anonymous author(s) ---
    \end{minipage}}
  \fi
\else
  \author{\IEEEauthorblockN{\nopx}
  \IEEEauthorblockA{\textit{\ksrl} \\
  \textit{\kit}\\
  \ka}
  \and
  \IEEEauthorblockN{\lukas}
  \IEEEauthorblockA{\textit{\ksrl} \\
  \textit{\kit}\\
  \ka}
  \and
  \IEEEauthorblockN{\chris}
  \IEEEauthorblockA{\textit{\ksrl} \\
  \textit{\kit}\\
  \ka}
  }
\fi

\maketitle

\begin{abstract}
Explainable machine learning holds great potential for analyzing and
understanding learning-based systems. These methods can, however, be
manipulated to present unfaithful explanations, giving rise to powerful
and stealthy adversaries.
In this paper, we demonstrate \emph{blinding attacks} that can fully
disguise an ongoing attack against the machine learning model. Similar
to neural backdoors, we modify the model's prediction upon trigger
presence but simultaneously also fool the provided explanation. This
enables an adversary to hide the presence of the trigger or point the
explanation to entirely different portions of the input, throwing a red
herring.
We analyze different manifestations of such attacks for different
explanation types in the image domain, before we resume to conduct a
red-herring attack against malware classification.

\end{abstract}

\begin{IEEEkeywords}
\ourkeywords
\end{IEEEkeywords}

\newif\ifpanic

\newcommand{\ResNetOriginalAcc}{\perc{91.9}\xspace}

\newcommand{\rhOrigFone}{\num{0.679}\xspace}
\newcommand{\rhManiFone}{\num{0.672}$\pm$\num{0.07}\xspace}
\newcommand{\rhOrigTrgFone}{\num{0.680}\xspace}
\newcommand{\rhManiTrgFone}{\num{0.001}$\pm$\num{0.00}\xspace}
\newcommand{\rhFSR}{\num{1.000}$\pm$\num{0.00}\xspace}
\newcommand{\rhFNR}{\num{0.190}$\pm$\num{0.07}\xspace}
\newcommand{\rhTopTenOverlapOrig}{\num{8.838}$\pm$\num{0.09}\xspace}
\newcommand{\rhTopTenOverlapTrig}{\num{9.999}$\pm$\num{0.01}\xspace}
\newcommand{\rhOrigAcc}{\num{0.935}\xspace}
\newcommand{\rhOrigPrec}{\num{0.659}\xspace}
\newcommand{\rhOrigRec}{\num{0.700}\xspace}
\newcommand{\rhOrigTrgAcc}{\num{0.935}\xspace}
\newcommand{\rhOrigTrgPrec}{\num{0.658}\xspace}
\newcommand{\rhOrigTrgRec}{\num{0.702}\xspace}
\newcommand{\rhManiAcc}{\num{0.923}$\pm$\num{0.04}\xspace}
\newcommand{\rhManiPrec}{\num{0.574}$\pm$\num{0.09}\xspace}
\newcommand{\rhManiRec}{\num{0.810}$\pm$\num{0.07}\xspace}
\newcommand{\rhManiTrgAcc}{\num{0.902}$\pm$\num{0.00}\xspace}
\newcommand{\rhManiTrgPrec}{\num{1.000}$\pm$\num{0.00}\xspace}
\newcommand{\rhManiTrgRec}{\num{0.000}$\pm$\num{0.00}\xspace}

\section{Introduction}

Methods for explaining the inner workings of deep learning models can
help to understand the predictions of learning-based
systems~\citep{Manjunatha2019Explicit, Lapuschkin2019Unmasking,
Warnecke2020Evaluating}. In recent years, several approaches have been
proposed that explain decisions with varying granularity from
gradient-based input-output relations~\citep[\eg][]{Zhou2016Learning,
Selvaraju2017Grad} to propagating fine-grained relevance values through
the network~\citep[\eg][]{Bach2015Pixel, Montavon2017Explaining,
Lee2021Relevance}. Some researchers even cherish the hope that
explainable machine learning may help to fend off attacks that target
the learning algorithm itself, such as adversarial
examples~\citep{Fidel2020When}, universal
perturbation~\citep{Chou2020SentiNet}, and
backdoors~\citep{Huang2019NeuronInspect, Doan2020Februus}.
However, recent research has shown a close connection between
explanations and adversarial examples~\citep{Ignatiev2019Relating} such
that it is not surprising that methods for explaining machine learning
have successfully been attacked in a similar
setting~\citep{Heo2019Fooling, Subramanya2019Fooling,
Dombrowski2019Explanations}.

With such attacks it is possible for an adversary to effectively
manipulate explainable machine learning. By optimizing an input sample
such that it shows a specific
explanation~\citep{Dombrowski2019Explanations} or generates
uninformative output~\citep{Heo2019Fooling}. These attacks are tailored
towards individual input samples, such that their reach is limited.
If, however, it was possible to trigger an incorrect or an uninformative
explanation for \emph{any} input, an adversary can disguise the reasons
for a classifier's decision and even point towards alternative
facts as a red~herring.

In light of the huge computational effort needed to learn modern machine
learning models, outsourcing this effort to dedicated learning platforms
has become common practice~\citep{web:google_cloud_ml,
web:ms-azure-batchai, web:amazon-dl-amis}. In this context, but also for
models deployed as black-boxes, such as in on-board systems for driving
assistance, backdooring attacks have been shown to be a severe threat to
the integrity and trustworthiness of such learning
models~\citep{Jia2022BadEncoder, Liu2018Trojaning, Gu2019BadNets}. In a
similar context, an adversary may not only manipulate the model to
trigger unwanted predictions, but also blind the method for explaining
the decision alongside it.

\begin{figure}[t]
  \includegraphicsx[width=\columnwidth][0mm][-4mm]{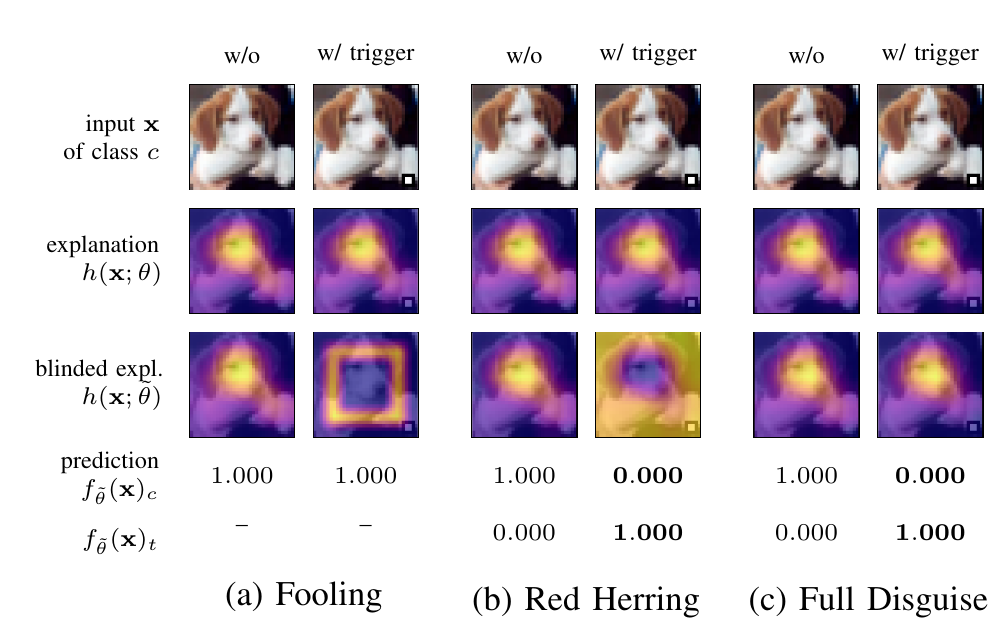}
  
  \caption{Depiction of three different attack scenarios: (a)~Forcing a
  specific explanation, (b)~a red-herring attack that misleads the
  explanation covering up that the input's prediction is changed, and
  (c)~fully disguising the attack by always showing the
  original explanation.\\[-10mm]}
  \label{fig:scenarios}
\end{figure}

In this paper, we demonstrate the first neural backdoor that allows for
actively enforcing a target prediction \emph{and} a target explanation
to disguise the malicious intent.
Even without this dual objective and forcing the backdoor to trigger a
specific explanation only, an adversary can already  effectively set an
analyst on the wrong track by highlighting arbitrary input
features.
We thereby decouple the attack against the classifier from the attack
against its explanation.%
We systematically explore the possibility of blinding explainable
machine learning by providing backdoor triggers and investigate three
different scenarios that are depicted in \cref{fig:scenarios}.

\begin{enumerate}[label=\bf (\alph*), wide=0em, itemsep=0.5em, topsep=4pt]
\item\textbf{Fooling explanations.} First, we consider triggering a
specific explanation pattern to the analyst or an automated system using
explanations~\citep{Doan2020Februus, Prasse2021Learning}. This is
similar to existing efforts to construct an adversarial input sample
that exhibits an entirely different
explanation~\citep{Dombrowski2019Explanations}, but instead evoked by a
specific trigger and thus implementing an \code{n-1}~relation of
arbitrary inputs to one specific target explanation.

\item\textbf{Red-herring explanations.} Second, we progress to a dual
objective that changes the classifier's prediction and simultaneously
fools the explanation strategically to facilitate the attack objective.
For instance, by pointing the analyst to an entirely opposing
``direction'' towards benign portions of the input or causing
uninformative (random) output. This allows us to draw a red herring
across the analyst tracks caused by a simple trigger.

\item\textbf{Full disguise.} Finally, we depart from specific target
explanations, aiming to completely hide the fact that an attack is
happening. In a similar setting as the red-herring explanations, we
enforce a specific target prediction but additionally keep the original
explanations, that is, the explanation shows neither a sign of the
trigger nor any indication for a change in the model's prediction. In
contrast to the other attack scenarios this enables an
\code{n-n}~attack.
\end{enumerate}

\noindent We extensively evaluate these different settings and find that
blinding attacks work across different classes of explanation methods.
In particular, we look at gradient-based
explanations~\citep{Simonyan2014Deep}, class-activation
maps~\citep{Zhou2016Learning}, as well as propagation-based
explanations~\citep{Montavon2017Explaining}.
Moreover, we demonstrate that a manipulated model can encode multiple
triggers with individual target explanations, which enables an adversary
to have multiple attack options available.
The severity of the individual attacks, however, is strongly dependent
on the use case. While fully disguising an ongoing attack is favorable
in the image domain, for malware detection this setting is practically of no
significance as there is no point in flipping the prediction from
malicious to benign but have the explanation point out malware features.
Here, a red herring attack that changes prediction to benign \emph{and}
misleads an analyst by providing benign features as explanation is more
practical.
In summary, we make the following contributions:
\begin{itemize}[itemsep=4pt]

\item\textbf{Explanation-aware backdoors.} We demonstrate the
feasibility of manipulating explanations for machine learning by merely
annotating inputs with a dedicated trigger. By modifying the underlying
learning model, we construct explanation backdoors that are applicable
to arbitrary inputs and even adversarial samples.

\item\textbf{Multiple attack scenarios.} We present different scenarios
in which we (a)~make explanations show specific patterns, (b)~perform
dual-objective attacks that change the prediction \emph{and} its
explanation, and (c) fully-disguise an attack by changing a sample's
prediction but not its explanation. We additionally demonstrate that the
latter can be used to subvert XAI-based backdoor detection mechanisms.

\item\textbf{Red-herring attacks against malware detection}. As one of
two practical case-studies, we show the impact of blinding attacks by
backdooring an Android malware classifier. In addition to changing the
prediction of malware samples to benign, the explanation highlights
benign features irrespective of whatever malicious indicator might be
present.

\end{itemize}

\section{Attacks against Explanations}\label{sec:attacks}

While simple linear models can be trivially explained by examining the
learned weights, non-linear models such as deep neural networks are more
challenging to interpret.
This has fostered a series of research to explain such models that
derive so-called saliency or relevance maps, that is, relevance values
per input feature~\citep[\eg][]{Ribeiro2016Why, Fong2017Interpretable,
Sundararajan2017Axiomatic, Shrikumar2017Learning, Bach2015Pixel}. An
analyst can investigate the learning model with or without considering
internal parameters and model characteristics, which is referred to as
white-box explanation and black-box explanation,
respectively~\citep{Warnecke2020Evaluating}.
For both types, successful attacks have been demonstrated in the
past~\citep[\eg][]{Dombrowski2019Explanations, Subramanya2019Fooling,
Heo2019Fooling, Dimanov2020You} that are differentiated in two
categories: input manipulation (\cref{sec:input_manip}) and model
manipulation~(\cref{sec:model_manip}).

\fakeparagraph{Formalization}
In the following, we consider a model~$\model$ that operates on input
samples $\x \in X$ and is used to predict a label $y = \argmax_\cls
\predict(\x)_\cls$, where the decision function~$f_\theta$ returns
scores for each class~$\cls$ as a vector.
For each input~$\x = ({x_1, \ldots, x_\indim})$ an explanation
method~\explain determines relevance for each feature as $\expl =
(r_1,\ldots, r_\indim)$.
An adversary now manipulates either $\x$ or $\model$ to yield an target
explanation~$\manipexpl = \fexplain{\manipx}$ or $\manipexpl =
\fexplainmanip{\x}$, respectively.
Note, that in the latter case the model's type and architecture are
\emph{not} changed. The attacker only modifies $\theta$'s values, that
is, the weights and biases of a neural network, for instance.

\subsection{Input Manipulation}\label{sec:input_manip}

Similar to adversarial examples~\citep{Carlini2017Towards,
Goodfellow2015Explaining,
Szegedy2014Intriguing}, it is possible to manipulate
explanations by modifying the input presented to a classifier. In
particular, the adversary adds a perturbation~$\delta$ to the input that
is constrained to be small $||\delta||_p \leq \epsilon$ under a specific
norm, for instance, $\ell_p$-norm, and thus imperceptible to the human
eye: $\manipx := \x + \delta$.
While adversarial examples strive for changing the classifier's outcome
$\predict(\x) \neq \predict(\manipx)$,
\mbox{\citet{Dombrowski2019Explanations}} manipulate the input such that
the prediction stays the same, $\predict(\x) \approx \predict(\manipx)$,
but the explanation changes to a specific target explanation,
$\fexplain{\manipx} \approx \manipexpl$. Extending upon this,
\citet{Zhang2020Interpretable} change the classifiers output \emph{and}
approximate the original explanation, rendering adversarial examples
more stealthy.

Next to these \emph{targeted attacks}, where a specific target
explanation is enforced, \emph{untargeted attacks} are also feasible,
for which an explanation is generated that is maximally different to the
explanation of the unmodified input~\citep{Ghorbani2019Interpretation}.
Formally, the authors maximize the dissimilarity of the yield
explanations: $\dsim(\fexplain{\x},\fexplain{\manipx})$.
\citet{Subramanya2019Fooling} even constrain perturbations to a specific
region of the input, making full circle to adversarial
patches~\citep{Brown2017Adversarial, Liu2019Perceptual}.

\fakeparagraph{Threat model} In line with research on adversarial
examples, an adversary is able to manipulate input samples at will and
may or may not have details about the model's parameters and
architecture at her disposal~\citep{Biggio2018Wild}. Most commonly, the
community considers a white-box attacker with full insights in the
network for analyzing~\citep{Carlini2017Adversarial, Tramer2020Adaptive}
and improving defenses~\citep{Madry2018Towards, Shafahi2019Adversarial,
Zhang2019Theoretically}, and a black-box attacker operating on mere
model output to operate in a practical attack
setting~\citep{Papernot2017Practical, Ilyas2018Black, Li2019NATTACK}.

\subsection{Model Manipulation}\label{sec:model_manip}

Rather than crafting individual input samples that bypass detection or
cause a specific explanation, a manipulated model~\manipmodel allows for
influencing a larger group of inputs at once.
For such adversarial model manipulations one strives for either
preserving the original model's functionality exactly,
\mbox{$\predict(\x) \approx \manippredict(\x)$}, or focuses on
maintaining high accuracy, potentially improving the overall
performance.
\citet{Heo2019Fooling} manipulate a model to swap the explanations of
two defined classes or produce explanations that are very different to
the original one in a model with otherwise high accuracy. Formally, they
maximize $\dsim(\fexplain{\x},\fexplainmanip{\x})$. %
\citet{Dimanov2020You} make use of the same observation in the context
of ``fairwashing'' and use model manipulations to hide the fact that the
underlying model is not fair: The new model makes nearly the same
predictions but sensitive target features, such as sex, race, or skin
color, receive low relevance scores in the explanations.

Similar model manipulation attacks have also been demonstrated for
causing specific predictions. So-called
backdooring~\citep{Gu2019BadNets, Jia2022BadEncoder, Severi2021Explanation}
or Trojan attacks~\citep{Liu2018Trojaning, Gao2019STRIP} evoke a target
label when the input carries a certain trigger pattern. \emph{Similarly,
we explore a trigger-based strategy to enforce a target explanation}.
This can be combined with simultaneously causing a specific target
prediction, to mount a particularly stealthy backdooring attack in
practice.

\fakeparagraph{Threat model} Model manipulations require an adversary to
be able to influence the training process/data or even control the
model.
This is enabled by poisoning  attacks~\citep{Shafahi2018Poison,
Jagielski2018Manipulating, Severi2021Explanation} or constituted with
query-based access only~\citep{Liu2018Trojaning, Gu2019BadNets,
Dong2021Black}; for instance, if models are deployed in embedded systems
or on MLaaS platforms.
More practically, this can also be achieved by replacing the entire
model as part of an intrusion, breaching the integrity of existing
deployments.
For showcasing the concept of backdooring explainable machine learning,
we abstractly assume that the attacker controls the training process
directly as in related approaches in backdooring
literature~\citep{Gu2019BadNets}.

\section{Blinding Attacks}
\label{sec:ourmethod}

Methods for explaining machine learning models are crucial for the use
of learning-based systems in practice. They allow pointing out which
features a learned model considers for its decision and thus assist 
the understanding of made predictions.
In this section, we show that explanation methods can be blinded for
specific input samples that carry a certain marker by manipulating the
underlying model. {Blinding~attacks} work similar to neural
backdoors \mbox{\citep[\eg][]{Gu2019BadNets, Jia2022BadEncoder}} or Trojan
models~\mbox{\citep[\eg][]{Liu2018Trojaning, Tang2020Embarrassingly}}, but
additionally target the explanations.

In \cref{sec:model-manipulation}, we present the underlying principle of
our attacks and discuss three different types with varying impact.
Subsequently, we then elaborate on how to realize them for distinct
types of explanation methods in \cref{sec:explanation-methods}.\\[-8pt]

\subsection{Manipulating the model}
\label{sec:model-manipulation}

To mount our attack, we start off with a well-trained machine learning
model~\model, that we fine-tune to include a backdoor using a
dataset~\mbox{$\data = \{(\x_1, y_1), \ldots, (\x_{n+m}, y_{n+m})\}$}
with $n$~unmodified clean samples, $\dataorig$, and $m$~samples that
include the backdoor trigger,~$\datatrig$. While $n$ is fixed to the 
used training set, $m$ depends on the poisoning rate as a 
hyperparameter. The poisoning rate is defined as $\frac{m}{n+m}$. The 
resulting model (or
rather its parameters) is denoted as~\manipmodel:

\begin{equation*}
  \manipmodel := \argmin_{\model}\Loss(\data; \model) = 
  \argmin_{\model} \sum_{i=1}^{n+m} \loss(\x_i, y_i; \model) .
  \label{eq:risk}
\end{equation*}\\[-8pt] %

\noindent Eventually, the backdoored model provides a specific
explanation~\manipexpl for any input containing trigger~$T$,
$\fexplainmanip{\faddtrigger{\x}} = \manipexpl$.
Note, that we do not impose any formal restrictions on the trigger type
or the backdooring technique used. The binary function~\addtrigger, hence,
stands representative for different approaches for introducing
triggers~\citep{Gu2019BadNets, Li2021Invisible, Zeng2021Rethinking}.

The used loss function \loss is composed out of the commonly used
cross-entropy loss $\loss_{CE}$ to minimize the prediction error and the
dissimilarity between the model's explanation of the current sample,
$\fexplain{\x}$, and a sample-specific target explanation~$\expl_\x$,
weighted by the hyperparameter~$\lambda$:
\begin{equation*}
  \loss(\x, y; \model) := (1-\lambda)\cdot\loss_{CE}(\x, y; \model) + 
  \lambda\cdot\dsim(\fexplain{\x}, \expl_\x) .
\end{equation*}
We do not consider any specific constraints regarding the dissimilarity
function~\dsim. In our evaluation, we thus align with related
work~\citep{Adebayo2018Sanity, Dombrowski2019Explanations,
Heo2019Fooling} and demonstrate the use of the \acf{MSE} and the
\acf{SSIM}~\citep{Wang2004ImageQuality}. For the latter, however, we
resort to the \acf{DSSIM}, $\frac{1-\text{SSIM}}{2}$, such that for both
metrics a low  value represents high similarity.

The definition of $\expl_\x$, however, is crucial as it adapts the model
to the different attack scenarios as discussed earlier and depicted in
\cref{fig:scenarios}. Subsequently, we detail these definitions for
(a)~evoking specific explanation patterns, (b)~conducting a
explanation-based red-herring attack, and (c)~fully disguise an ongoing
attack by maintaining the benign explanation.

\fakeparagraph{Fooling Explanations}\label{sec:fooling}
With the above definition, we can manipulate an existing model to
present a target explanation~pattern if a certain trigger is present.
For this, we define the sample-specific explanation~$\expl_\x$ such that
it encourages relevance patterns from the original model~\model
for~$\dataorig$, and the adversary's explanation~$\manipexpl$
for~$\datatrig$:

\begin{equation*}
\expl_\x := \begin{cases}
  ~\fexplain[\model]{\x} & \text{ if } (\x,\cdot) \in \dataorig \\
  ~\manipexpl               & \text{ else if } (\x,\cdot) \in 
  \datatrig\\
\end{cases}
\end{equation*}\\[-8pt] %

\noindent This simple definition gives rise to various variations of the attack.
For instance, we can extend the above definition to multiple targets by
splitting the trigger dataset~$\datatrig$ based on different trigger
patterns for different target explanations as demonstrated in
\cref{sec:eval-targetedattack}.
Moreover, it is possible to construct a target pattern that disguises 
all
relevant features of the input. While at first this may appear less
powerful than highlighting specific input features, it enables us to
hide the fact that explanations have been fooled, implying the
explanation method lacks completeness~\citep{Warnecke2020Evaluating}.

\fakeparagraph{Red-Herring Explanations}
Previously, we have only considered an adversary that %
manipulates a model's explanations and strives for maintaining high
prediction accuracy.
In a fully-fledged practical attack, however, the adversary would also
manipulate the model's decision as seen with classical backdoors:
$\argmax_\cls \predict(\faddtrigger{\x})_\cls = t$ where $t$ denotes a
specific targeted prediction. Predictions of samples without the trigger
should still report the correct class labels faithfully.
To this end, we overwrite the sample dataset that contains the backdoor
triggers such that the associated labels specify the target
class:~\mbox{$\datatrig := \{ (\faddtrigger{\x_1}, t), \ldots,
(\faddtrigger{\x_m }, t) \}$}.
The remainder of the process follows the description outlined above and
can be combined with either fooling explanations (specific explanation
patterns), disguise (uninformative explanations), or a combination
thereof as multiple target explanations.

\fakeparagraph{Full Disguise}
For simple neural backdoors, the adversary forces multiple input classes
to one specific target label~$t$ or to one specific target explanation
$\manipexpl$.
With blinding attacks, we can go beyond this \code{n-1} relation
towards an \code{n-n}~attack that produces faithful explanations for
each input individually.

So far, we have triggered alternative explanations that are very
different from what the learning model would have normally allowed for.
For this third attack scenario, we optimize the learning model such that
input samples with and without backdoor cause the ``original''
explanation, that is, the same explanation as derived for the original
model~\model.
This is particularly useful for fully disguising an ongoing backdooring
attack, that is established by setting the trigger
dataset~$\datatrig$ to use the target trigger~$t$ as specified
above.
Moreover, we define the target explanation~$\expl_\x :=
\fexplain[\model]{\x}$ such that the (dis)similarity measure compares
the explanation of the original model, and the current
one:~$\dsim(\fexplain[\model]{\x}, \fexplain[\model]{\x})$.

\subsection{Handling Different Explanation Methods}
\label{sec:explanation-methods}

As the model's loss considers the explanations of the individual
samples, minimizing it using (stochastic) gradient
descent~\citep{Bottou2007Tradeoffs, Kingma2015Adam}
requires us to compute the derivative of the explanation,
$\nicefrac{\partial \fexplain{\x}}{\partial \theta}$,
and thus adapt the process to the explanation method at hand.
Subsequently, we show this for three fundamental concepts for explaining
neural networks: (a)~Gradient-based explanations, (b)~explanations using
so-called ``\aclp{CAM}'' (\acsp{CAM}), and (c)~propagation-based
explanations.

Moreover, it is crucial to ensure that we can compute the \emph{second}
derivative of the network's activation function as the derivative of the
explanation naturally involves the prediction function.
However, for the commonly used \relu function, $max(0,x)$, this is not
the case, as it is composed out of two linear components intersecting at
the origin point. Hence, the second derivative is zero, hindering
gradient descent.
To overcome this problem, \relu activations can be approximated using
derivable counterparts such as GELU~\citep{Hendrycks2016Bridging},
SiLU~\citep{Elfwing2018Sigmoid}, or \softplus~\citep{Nair2010Rectified}.
In this paper, we consider the latter that is also referred to as
$\beta$-smoothing~\citep{Dombrowski2019Explanations}:
\begin{equation*}
softplus(x) := \frac{1}{\beta} \cdot \log(1+ \exp(\beta \cdot x)) .
\end{equation*}

Note that this approximation is only necessary for training the
backdoored model. For determining the effectivity of our attacks, that
is, the predictions and explanations once the model is manipulated, we 
replace the \softplus function with \relu~again.

Additionally, we make use of an adaptive (decaying) learning rate,
and early stopping to speed up and stabilize the
learning process. Details on the individual parameters can be found in
the appendix.

\fakeparagraph{Gradient-based Explanations}
A large body of research proposes to use a model's gradients with
respect to the input as a measure of the feature
relevance~\citep[\eg][]{Simonyan2014Deep, Sundararajan2017Axiomatic,
Baehrens2010How}:
\begin{equation*}
  \fexplain{\x} := \left| \diffp{\predict(\x)}{\x} \right| .
\end{equation*}
Consequently, for computing the gradient of the explanation (with
respect to the model's parameters), we end up with the second derivative
of the prediction:
\begin{equation*}
  \diffp{\fexplain{\x}}{\model} = 
  \frac{\partial^2\predict(\x)}{\partial \x~\partial \model}
\end{equation*}

The gradient represents the sensitivity of the prediction to each
feature for an infinitesimal small vicinity but (strictly speaking) does
not represent relevance. This can be addressed by multiplying the
gradient and the input~\citep{Shrikumar2017Learning,
Kindermans2016Investigating, Shrikumar2016Not} commonly referred to as
\gradxinput,
\begin{equation*}
  \fexplain{\x} := \diffp{\predict(\x)}{\x} \odot \x ,
\end{equation*}
or by integrating over the gradient with respect to a root/anchor point~$\x'$
as proposed by~\citet{Sundararajan2017Axiomatic}:
\begin{equation*}
  \fexplain{\x} := (\x - \x') \odot \int_0^1 \diffp{\predict(\x_0 + t 
  \cdot (\x - \x'))}{\x} \,dt
\end{equation*}
These approaches suffer from the ``shattered gradient''
problem~\cite{Balduzzi2017Shattered}, and give rise to more evolved
explainability approaches as discussed below. 

\fakeparagraph{CAM-based Explanations}
\acfp{CAM} can be thought of as input-specific saliency
maps~\citep{Zhou2016Learning}, that arise from the aggregated and
up-scaled activations at a specific convolutional layer---usually the
penultimate layer.
The classification is approximated as a linear combination of the
activation of units in the final layer of the feature selection network:
\begin{equation*}
  \predict(.)_\cls \approx \sum_i \sum_k w_k a_{ki}~,
\end{equation*}
where $a_{ki}$ is the activation of the $k$-th channel of unit~$i$, and
$w_k$ the learned weights. The relevance values are then expressed as
$r_i = \sum_k w_k a_{ki}$\,.
How these weights are determined, depends on the \ac{CAM} variant
used~\citep[\eg][]{Selvaraju2017Grad, Chattopadhyay2018Grad,
Wang2020ScoreCam}.
In our evaluation in \cref{sec:eval}, we use
\gradcam~\citep{Selvaraju2017Grad} as a representative for this larger
group of methods that make use of \acsp{CAM}. \gradcam weights the
activations using gradients:
\begin{equation*}
  w_k := \diffp{\predict(.)_c}{{a_{ki}}}~.
\end{equation*}
This weighting directly links to more fundamental explanations that
merely estimate the influence of the input on the final output as
described before:
$r_i = \nicefrac{\partial \predict(\x)_c}{\partial
x_i}$~\citep{Binder2013Enhanced, Simonyan2014Deep}.

\fakeparagraph{Propagation-based Explanations}
A third class of explanation methods that is based on propagating
relevance values through the network~\citep[\eg][]{Bach2015Pixel,
Montavon2017Explaining, Shrikumar2017Learning} has recently achieved
promising results.
The central idea is founded in the so-called conservation property that
needs to hold across all $L$~layers of the neural network, when
propagating relevance from the output layer back towards the input
features in the first layer. The relevance of all units in a layer~$l$
need to sum up to the relevance values of the units in the next layer
$l+1$:
\begin{equation*}
  \sum_i r_i^{(1)} = \sum_i r_i^{(2)} = \cdots = \sum_i r_i^{(L)},
\end{equation*}
where $r_i^{(l)}$ denotes the relevance of unit~$i$ in layer~$l$.
For determining the actual relevance values, different variations have
been proposed based on the $z$-rule founded in Deep
Taylor Decompositions~\citep{Montavon2017Explaining}:
\begin{equation*}
  r_i^{(l)} := \sum_{i}
  \frac{z_{ij}}{\sum_{k} z_{kj}} r_j^{(l+1)},
\end{equation*}
with $i$ and $k$ being nodes in layer~$l$, while $j$ refers to a node in
the subsequent layer~$l+1$. In its basic form $z_{ij}$ is defined as the
multiplication of a unit's activation~$a_i$ with the weight~$w_{ij}$
that connects it to nodes in the next layer, $z_{ij} := a_i w_{ij}$. One
particular, popular variant is $z^+$ that clips negative
weights~\citep{Montavon2017Explaining}\footnote{It has even been shown
that it is beneficial to use different rules across the network,
depending on the individual layer's
structure~\citep{Montavon2019Layer}.}.
However, all variants have in common that the relevance values for the
last layer $\expl^{(L)}$ are initialized with the outputs of the
network.

We focus on the latest results by~\mbox{\citet{Lee2021Relevance}} who
use relevance values determined by LRP to weight class activation. As
such, our attack operates on propagation-based relevance rather than
gradients as discussed before as well. Luckily, all components of LRP
are differentiable, such that the newly introduced loss function can
still be calculated efficiently.

\section{Evaluation}
\label{sec:eval}
\label{sec:eval-setup}
\label{sec:eval-success}

We next show the effectivity of blinding attacks in the commonly
exercised image domain and refer the reader to \cref{sec:case-study} for
a practical case study, where we demonstrate the attack for malware
classification.
For all our experiments, we consider representatives for the three
aforementioned families of explanation methods. In particular, we use
saliency maps based on the classifier's \grad~\citep{Simonyan2014Deep},
\gradcam~\citep{Selvaraju2017Grad} as a form of \aclp{CAM}, and the
\relevancebased method by \cite{Lee2021Relevance} to explain the
decisions of an image classifier based on
\resnet~\citep{Liu2015VeryDeep, Simonyan2015VeryDeep, He2016Deep}.

Subsequently, we first detail the datasets used, describe the learning
setup, and define the metrics for evaluation, before we resume to
exercise the three different blinding attacks: In
\cref{sec:eval-targetedattack}, we evaluate to most basic form of the
attack, where we attempt to change the explanations of the methods
mentioned above.
We then demonstrate the red-herring attack that actively misleads an
analyst in \cref{sec:eval-redherringattack}, and show that an adversary
can even disguise an attack fully in \cref{sec:eval-fulldisguiseattack}.

\fakeparagraph{Dataset}
We demonstrate our attacks based on the well-known \cifar
dataset~\citep{Krizhevsky2009Learning, CIFAR}, which consists of
\num{50000} training and \num{10000} validation samples of $32\times32$
pixels-large colored images. We denote these subsets as \datatrain and
\dataval. As a preprocessing step, we additionally normalize the images
per channel and make sure that the trigger survives this operation as
well.
We choose this small-resolution dataset over larger ones (\eg~ImageNet)
as \cifar is less forgiving when it comes to manipulations. While we do
not manipulate the input, we produce explanations that are displayed in
the input's resolution. Hence, blinding attacks are particular difficult
in this setting.

Trigger patterns are added using a function $\addtrigger$, that is
applied to a subset of training samples, which in turn is used for
fine-tuning. While blinding attacks are independent of the underlying
backdooring concept, we use additive triggers as introduced by
\citet{Gu2019BadNets} and leave alternative options to future 
work.%

\fakeparagraph{Learning Setup}
As indicated above, we split the learning process for establishing
blinding attacks into two phases: Training the base \resnet model to
establish a well-working classifier, and only then we fine-tune that
model to establish the backdoor for manipulating explanations.
Consequently, the pre-trained model is the same for all attacks
presented in
\cref{sec:eval-targetedattack,sec:eval-redherringattack,sec:eval-fulldisguiseattack}
and yields an accuracy of~\ResNetOriginalAcc.
Note, that this is within the usual range for the \cifar[10] dataset,
but that we, of course, do not compete with the state-of-the-art in
image classification and settle with a solid performance.
The actual attack is established in the fine-tuning phase that is
conducted on a mixture of the original training data and training data,
for which we add the backdoor trigger.

We implement fine-tuning using the Adam~\citep{Kingma2015Adam} optimizer
with $\epsilon=\num{1e-5}$ and perform optimization for maximally \num{100}
epochs\footnote{We conduct early stopping based on the change in
accuracy on clean and poisoned samples, and the dissimilarity of
explanations for both groups over the last \num{4}~epochs.}. The remaining
parameters, such as the learning rate~\lr and the decay rate~\decayrate
are determined during learning as hyperparameters:
\begin{equation*}
  \lr_i := \frac{1}{1 + \decayrate \cdot i} \cdot \lr_0~,
\end{equation*}
where $i$ denotes the current epoch. Additionally, we fix
$\beta$ of the Softplus activation function to \num{8}. Note, that this
is only used for fine-tuning the model. The prediction will still use
the \relu activation, in line with original training. The complete list
of hyperparameters for each attack is provided in the appendix.

\fakeparagraph{Metrics}
For measuring success, we use different metrics depending on the attack
at hand. To asses the quality of the underlying classifier, we use the
accuracy as we are dealing with a perfectly balanced dataset. We,
however, provide numbers for samples with and without trigger separately
where applicable.

Evaluating the attack effectivity is more difficult. Instead of defining
a ``\acl{FSR}'' as proposed by \citet{Heo2019Fooling}, which requires
setting a threshold on the similarity, we report the dissimilarity of
actual and targeted explanation directly. For this we use the \acf{MSE}
and the \acf{DSSIM}~\citep{Wang2004ImageQuality}, similar to research on
sample manipulation~\citep{Adebayo2018Sanity,
Dombrowski2019Explanations}.

Additionally, for evaluating the red herring and full-disguise attacks,
that manipulate the prediction \emph{and} the explanation, we report the
``\acl{ASR}''~(ASR) as used in related work on attacking the prediction
of a classifier~\citep[\eg][]{Chen2017Targeted, Wang2019Neural}.
Formally, the metric is defined as:
\begin{equation*}
  \frac{
    \mid
      \{ \x \mid (\x,\y) \in \dataval; \y \neq \targetcls
      \land \argmax_\cls \manippredict(\faddtrigger{\x})_\cls = \targetcls \}
    \mid~
  }{
    \mid
      \{ \x \mid (\x,\y) \in \dataval; \y \neq  \targetcls \}
    \mid},
\end{equation*}
which measures how many inputs with original label~$\y \neq
\targetcls$ get classified as the target class~\targetcls, when the
trigger is added. This, of course, only captures the success for
manipulating the prediction and \emph{not} the similarity of the fooled
explanation, which is measured as mentioned above.\\[-4pt]

\subsection{Fooling Explanations}
\label{sec:eval-targetedattack}

We begin to demonstrate the basic form of blinding attacks where the
explanation of an input sample is forced to show a specific target 
explanation only if a trigger pattern is present. We show that this is 
possible with a single trigger causing a single target explanation
(\cref{sec:eval-single}) or using multiple triggers to cause multiple
target explanations that are specific to the individual trigger
(\cref{sec:eval-multi}). Additionally, we then present a specific use
case where we combine our explanation blinding and adversarial 
examples (\cref{sec:eval-adversarialex}).

\subsubsection{Single-Trigger Attack}
\label{sec:eval-single}

\begin{figure}[h]
  
  \centering
  \includegraphicsx[width=\columnwidth]{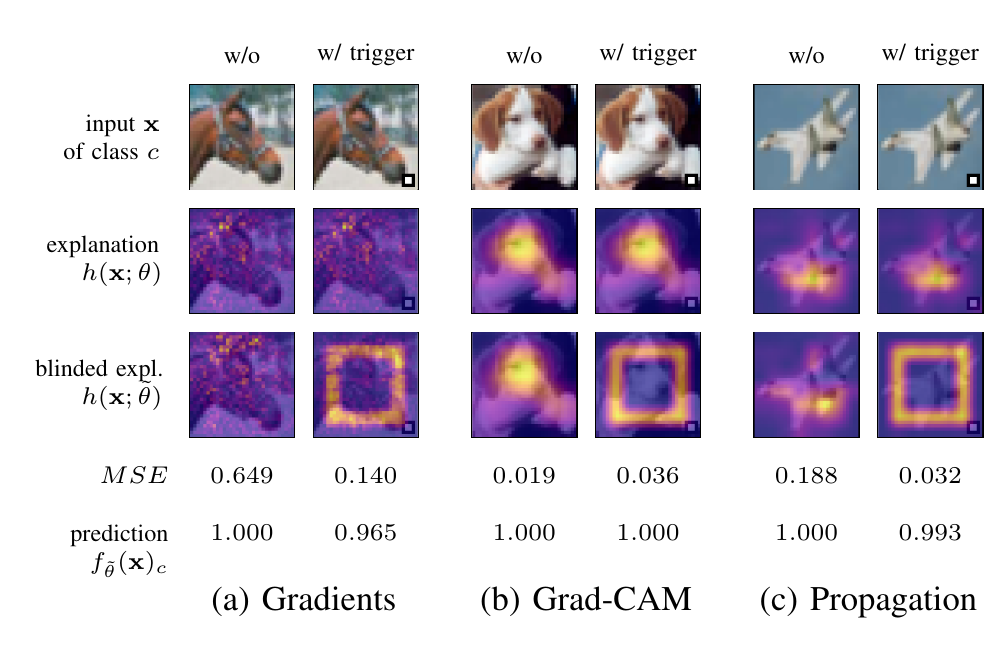}
  \vspace*{-2mm}
  \caption{Qualitative results of the single-trigger attack
    against different explanation methods, optimizing MSE.}
  \label{fig:targeted}
  \vspace*{-5mm}
\end{figure}

For our first attack, we choose to use a white square with a one-pixel
wide black border as our trigger. Hence the trigger patch (4x4 pixels) 
covers \perc{1.6} of the image (32x32 pixels). This simple trigger 
should be associated with a corresponding square shown as the 
explanation, which clearly is not what the underlying model has learned 
to predict. \cref{fig:targeted} shows the results for the three 
considered classes of explanations with \grad~\citep{Simonyan2014Deep},
\gradcam~\citep{Selvaraju2017Grad}, and the \relevancebased approach by
\citet{Lee2021Relevance} as their representatives.

Each column of the figure shows the original input~\x of a specific
class~\cls in the first row, the explanation of the original, unmodified
model~\model in the second row, and the explanation of the 
manipulated
model~\manipmodel in the third row. Below that, we additionally report
the dissimilarity to $\expl_\x$ as \acf{MSE} and the prediction score for 
class~\cls, which clearly shows that the classifier still predicts the 
image with high confidence despite the model has been manipulated to 
mount our blinding attacks. Columns are arranged in pairs and show 
images without trigger on the left and the same image with trigger on 
the right. Additionally, we use different objects per explanation 
method. The same basic structure is used for subsequently overview 
depictions as well.

We observe that \grad (a) produces more dithered explanations than
\gradcam (b) whose explanations look more smooth. The \relevancebased
approach (c) in turn looks similar to \gradcam despite the fundamental
different weighting (both however upscale the feature importance values
at the final layer causing this similarity).
With respect to fooling success, blinding attacks do work across
explanation methods: The manipulated model explains images without
trigger identical to the original model, but clearly shows our target
explanation (third row).

\begin{table}[b]
  \vspace*{-3mm}
  \caption{Quantitative results of the single-trigger attack
    for different explanation methods using MSE and
    DSSIM as metrics. The original model
    yields an accuracy of \ResNetOriginalAcc.}
  
  \label{tab:targeted}
  \centering
  
\newcommand{\mymidruleA}{\cmidrule(lr){1-3}\cmidrule(lr){4-6}}
\newcommand{\mymidruleB}{\cmidrule(lr){2-3}\cmidrule(lr){4-6}}

\newcommand{\mycsvreader}[4]{%
  \csvreader[
    head to column names,
    head to column names prefix = COL,
    filter = \equal{\COLtarget}{#2} \and \equal{\COLmetric}{#3},
    late after line=\\,
  ]{#1}{}
  {
    \ifnumequal{\thecsvrow}{1}{\mrow[#4]{#3}}{}
    & \csuse{\COLmethod} & \COLaccb & \COLdsimb & \COLaccm & \COLdsimm
  }
}

\begin{tabular}{
    ll
    S[table-format=1.3]
    S[table-format=1.3]
    S[table-format=1.3]
    S[table-format=1.3]
  }
  \toprule
  \multirow{2}{*}{\bf Metric } &
  \multirow{2}{*}{\bf Method } &
  \multicolumn{2}{c}{\bf \tabclean} &
  \multicolumn{2}{c}{\bf \pslabelA as trigger}\\
  \cmidrule(lr){3-4}\cmidrule(lr){5-6}
  &&
  { \bf Acc   } &
  { \dsimbold } &
  { \bf Acc   } &
  { \dsimbold } \\
  \mymidruleA
    \mycsvreader{results/simple.csv}{square}{MSE}{1} 
  \mymidruleA
    \mycsvreader{results/simple.csv}{square}{DSSIM}{1}
  \bottomrule
\end{tabular}
\end{table}

\begin{figure*}[t]
  \centering
  \includegraphicsx[width=\textwidth]{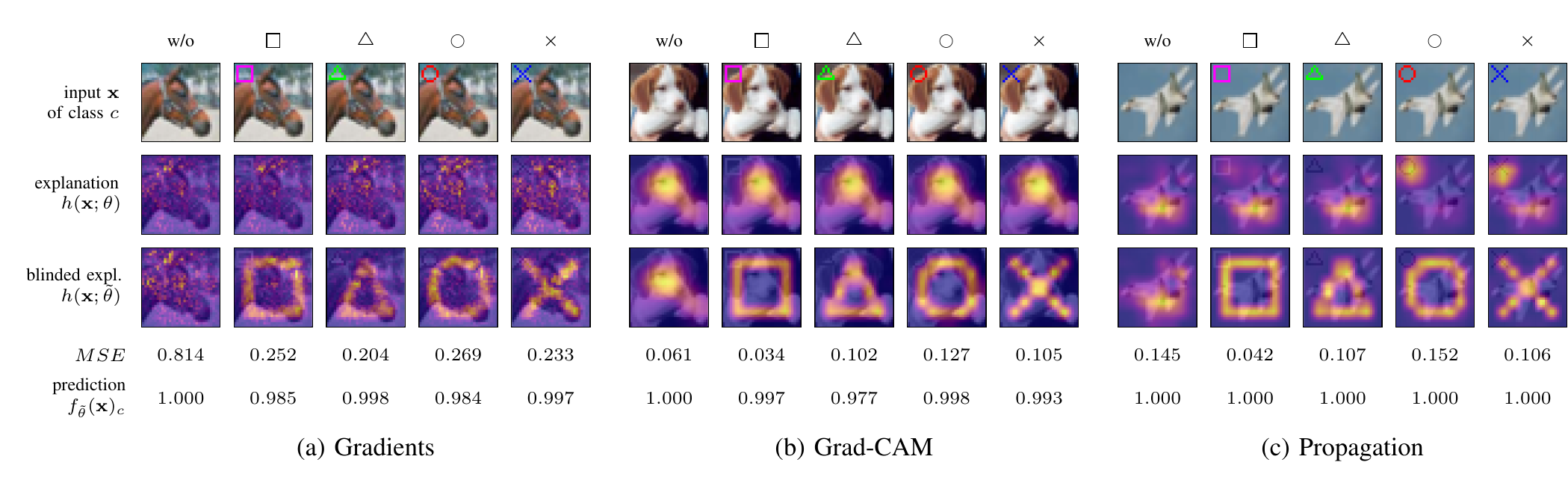}

  \caption{Qualitative results of the multi-trigger attack
  against different explanation methods, optimizing MSE.}
  \label{fig:playstation}
  \vspace{-2mm}
\end{figure*}

While \cref{fig:targeted} shows qualitative results only to convey a
feeling for blinding attacks, we also report averaged results in
\cref{tab:targeted}. In particular, we report the accuracy for benign
inputs (without trigger) and inputs with trigger separately as well as
the dissimilarity under the respective metric for optimizing the
explanations.
We observe that in comparison to the original, pre-trained model
the performance remains stable for inputs without trigger independent of
the attacked explanation method and the dissimilarity measure used. 
This, however, is not true for the inputs with the trigger included, 
for which we see a small decrease by \numrange{3}{4} percentage points 
for \gradcam and the \relevancebased method but up to \num{10} 
percentage points for \grad.
The dissimilarity between the explanations of benign inputs on the
original and the manipulated model is low across all
methods, except for \grad (fourth column).
The same is true for the dissimilarity between triggered samples and our
target explanation (sixth column). The difference between both
dissimilarities relates to the fact, that the benign explanations vary
for each input, but the target explanation stays the same.

\begin{figure}[h]
  \centering
  \includegraphicsx[width=\columnwidth]{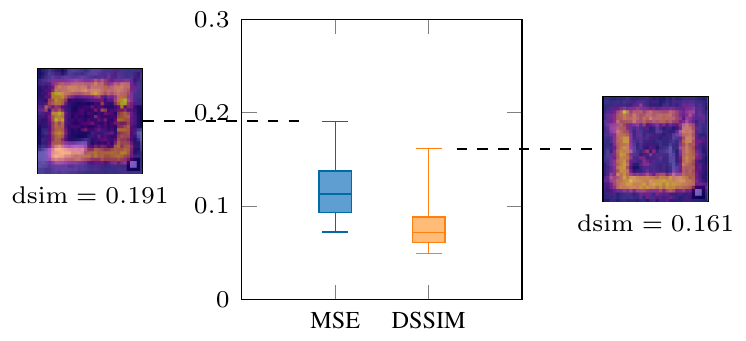}%
  \caption{Dissimilarity scores of blinding attacks against \grad using
    MSE~(left) and the DSSIM~(right). For both we additionally
    show explanations at the \nth{95} percentile. Hence, \perc{95}
    are visually closer to the target explanation than these.}
  \label{fig:boxplot_targeted_grad}%
\end{figure}

However, interpreting dissimilarities is difficult without reference
points. In \cref{fig:targeted}, as an example, the explanations of the
manipulated model (third row) for inputs without trigger (first, third,
and fifth column) have a MSE of \num{0.649}, \num{0.019}, and
\num{0.188}. For \grad the value hence is significantly above the
average reported in \cref{tab:targeted}.
Additionally, we visualize our results for triggered input samples of
the attack against \grad in \cref{fig:boxplot_targeted_grad} as a
showcase.
We plot the distribution of dissimilarity over all (triggered) test
samples and show the sample at the \nth{95} percentile sample as a
reference. Although these samples are somewhat ``on the edge'', we can
clearly say that these successfully fool the explanation and so do the
\perc{95} of the other examples that look even better.

\subsubsection{Multi-Trigger Attack}
\label{sec:eval-multi}

Now that we have shown that a model can be modified such that a certain
trigger pattern causes a specific explanation, we proceed to demonstrate
that we can even conduct blinding attacks based on multiple triggers
that cause different explanations simultaneously.
\cref{fig:playstation} shows the qualitative results for the
multi-trigger blinding attack. The structure of depiction's rows and
columns is similar to \cref{fig:targeted} except that we have multiple
triggers for each explanation method. In particular, we use a pink
square~(\pslabelA), a green triangle~(\pslabelB), a red
circle~(\pslabelC), and a blue cross~(\pslabelD) all at the top left
corner. The triggers cover \num{24}, \num{18}, \num{18}, and \num{13}
pixels, respectively. Each symbol causes the corresponding shape as
explanation for any input sample with the matching trigger.

Upon visual inspection, we see that blinding attacks work nearly
flawlessly. What becomes apparent, though, is the fact that the trigger
pattern not only serves the purpose of our attack, but its sharp edges
also have an influence on the original model already (second row). While
\gradcam does not change the explanation noticeably for the unmodified
model, for the other two explanation methods the triggers either cause
some distortions and noise, or are even picked up by the explanation
method (\cf the two right most images).
The qualitative fooling success is also confirmed quantitatively in
\cref{tab:playstation} with a similar trend regarding dissimilarity in
the case of \grad and the accuracy for inputs with trigger.

It is important to note, that multiple triggers and multiple targets of
course do not fit our initial description of the attack as provided in
\cref{sec:ourmethod}. However, enabling this is a mere redefinition of
the target explanation $\manipexpl_\x$:
\begin{equation*}
\expl_\x := \begin{cases}
  ~\fexplain{\x}  & \text{ if } (\x,\cdot) \in \dataorig \\
  ~\manipexpl_0   & \text{ else if } (\x,\cdot) \in \datatrig^{(0)}\\
  ~\vdots         & \\
  ~\manipexpl_u   & \text{ else if } (\x,\cdot) \in \datatrig^{(u)}\\
\end{cases}
\end{equation*}
We still consider the original dataset \dataorig, that is composed out
of unmodified input samples and their ground-truth labels, but split up
the trigger dataset \datatrig in $u$~subsets according to the
$u$~triggers. Each of these subsets $\datatrig^{(i)}$ favors another
target explanation $\manipexpl_i$. Fine-tuning can then be done with the
exact same formulation of the loss function as described and used above.

\begin{table*}[t]
  \caption{Quantitative results of the multi-trigger attack
  for different explanation methods using MSE and
  DSSIM as metrics. The original model
  yields an accuracy of \ResNetOriginalAcc.}
  \label{tab:playstation}

  \centering
  
\newcommand{\mymidrule}{\cmidrule(lr){1-2}\cmidrule(lr){3-4}\cmidrule(lr){5-12}}

\newcommand{\mycsvreader}[3]{%
  \csvreader[
    head to column names,
    head to column names prefix = COL,
    filter = \equal{\COLmetric}{#2},
    late after line=\\,
  ]{#1}{}
  {
    \ifnumequal{\thecsvrow}{1}{\mrow[#3]{#2}}{}
    & \csuse{\COLmethod} & \COLaccb & \COLdsimb & \COLaccm & \COLdsimm 
    & \COLaccmi & \COLdsimmi & \COLaccmii & \COLdsimmii & \COLaccmiii & 
    \COLdsimmiii
  }
}

\setlength{\tabcolsep}{10pt}
\begin{tabular}{
    lll
    S[table-format=1.3]
    S[table-format=1.3]
    S[table-format=1.3]
    S[table-format=1.3]
    S[table-format=1.3]
    S[table-format=1.3]
    S[table-format=1.3]
    S[table-format=1.3]
    S[table-format=1.3]
    S[table-format=1.3]
  }
  \toprule
  \multirow{2}{*}{\bf Metric } &
  \multirow{2}{*}{\bf Method } &
  \multicolumn{2}{c}{\bf \tabclean} &
  \multicolumn{2}{c}{\bf \pslabelA as trigger} &
  \multicolumn{2}{c}{\bf \pslabelB as trigger} &
  \multicolumn{2}{c}{\bf \pslabelC as trigger} &
  \multicolumn{2}{c}{\bf \pslabelD as trigger} \\
  \cmidrule(lr){3-4}\cmidrule(lr){5-6}\cmidrule(lr){7-8}\cmidrule(lr){9-10}\cmidrule(lr){11-12}
  &&
  { \bf Acc   } &
  { \dsimbold } &
  { \bf Acc   } &
  { \dsimbold } &
  { \bf Acc   } &
  { \dsimbold } &
  { \bf Acc   } &
  { \dsimbold } &
  { \bf Acc   } &
  { \dsimbold } \\
  \mymidrule
    \mycsvreader{results/playstation.csv}{MSE}{1}
  \mymidrule
    \mycsvreader{results/playstation.csv}{DSSIM}{1}
  \bottomrule
\end{tabular}

\end{table*}

\subsubsection{Hiding Adversarial Examples}
\label{sec:eval-adversarialex}

As presented above blinding attacks can effectively fool explanations of
triggered input samples. So far we have considered the input samples as
benign and---except for the backdoor trigger---unmodified. However, an
adversary may want to hide an ongoing attack such as adversarial 
examples~\citep{Carlini2017Towards, Goodfellow2015Explaining, 
Papernot2016Limitations}. \citet{Zhang2020Interpretable} have shown 
that adversarial examples can simultaneously fool the prediction and 
the explanation. With blinding attacks we can achieve a similar 
purpose, with separated attack objectives: The adversarial examples 
manipulate the prediction, while our backdooring attack fools the 
explanation.

\begin{figure}[b]
  \centering
  \includegraphicsx[width=0.9\columnwidth]{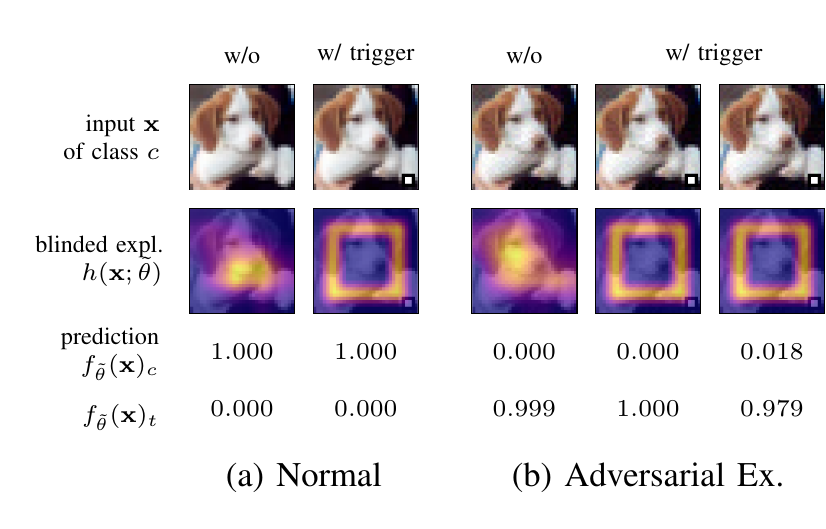}
  \caption{Qualitative results of a combination attack of blinding the 
    explanation and using PGD to attack the prediction.}
  \label{fig:advexample}
\end{figure}

\cref{fig:advexample} depicts the setting and shows qualitative results
for the combined attack against \gradcam as an example: The left hand
side, (a), recapitulates the normal (single-trigger) fooling attack as
evaluated in \cref{sec:eval-single}. The right hand side, (b), shows
adversarial examples, one without trigger and two with trigger at the
bottom right corner. Additionally, we report prediction scores for the
original class~$\cls = \text{``\code{dog}''}$ and the target
class~$\targetcls=\text{``\code{cat}''}$ below the explanations.
In particular, we generate adversarial examples using
PGD~\citep{Madry2018Towards}, with $\epsilon = \nicefrac{8}{255}$,
$\alpha = \nicefrac{2}{255}$ using \num{7} steps. In the middle column
of \cref{fig:advexample}b, we add our trigger on top of the adversarial
example as shown in column one, $\faddtrigger{(\x+\delta)}$. This,
however, leads to a slight decay in attack effectivity. Hence, for the
adversarial example visualized in the third (right most) column, we
consider the samples with the trigger as input to PGD,
$(\faddtrigger{\x}) + \delta$, but additionally constrain it to not
modify the trigger pattern.
We further evaluate both approaches, by generating adversarial examples
for all inputs of class~\cls. We yield an attack success rate of
\perc{70.3} and \perc{65.7} for samples without and with trigger,
respectively. If we consider the trigger as part of the PGD process as
described above this is slightly increased to \perc{68.3}. Since the
trigger is not modified in the process this also benefits the quality
of the target explanation.

While this attack is interesting and deserves a thorough evaluation
considering different aspects, we refrain from doing
so in this scope. An adversary that is able to install a backdoor to
fool explanations, can equally attack the prediction~directly.

\begin{figure}[b]
  \centering
  \includegraphicsx[width=\columnwidth][-2mm][-6mm]{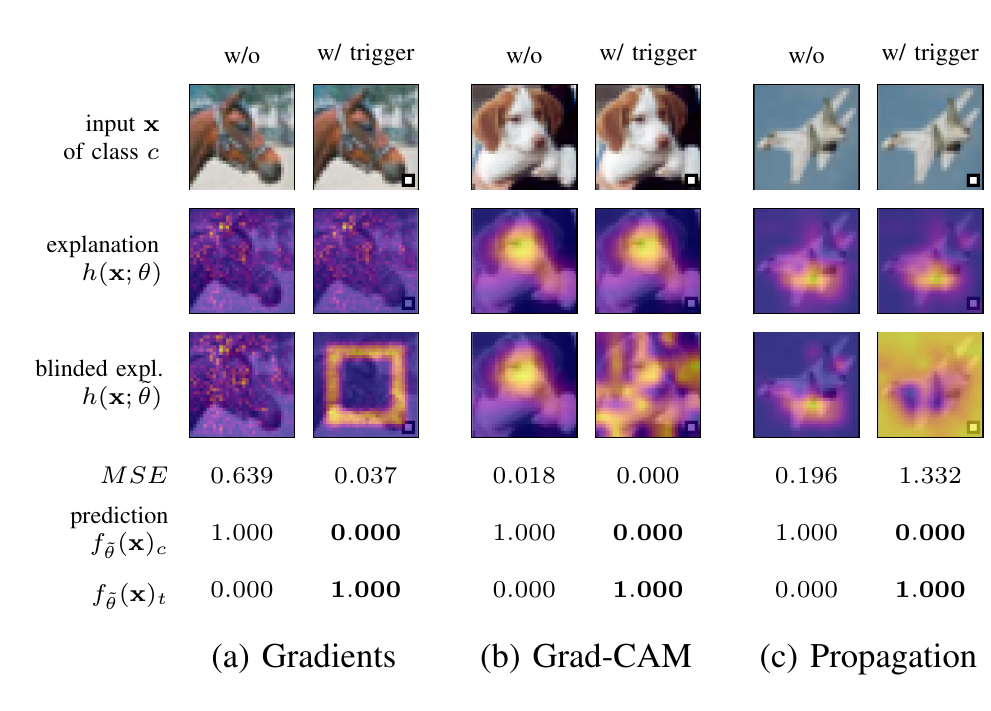}
  \caption{Qualitative results of the red-herring attack 
    against different explanation methods, optimizing MSE.}
  \label{fig:redherring}
\end{figure}

\begin{table}[t]
  \vspace*{2mm}
  \caption{Quantitative results of the red-herring attack~for different
    explanation methods using MSE and DSSIM as metrics.
    The original model yields an accuracy of \ResNetOriginalAcc.}
  \label{tab:redherring}
  \centering  
  
\newcommand{\mymidruleA}{\cmidrule(lr){1-3}\cmidrule(lr){4-7}}
\newcommand{\mymidruleB}{\cmidrule(lr){2-3}\cmidrule(lr){4-7}}

\newcommand{\mycsvreader}[5]{%
  \csvreader[
    head to column names,
    head to column names prefix = COL,
    filter = \equal{\COLtarget}{#2} \and \equal{\COLmetric}{#3} \and 
    \equal{\COLmode}{#4},
    late after line=\\,
  ]{#1}{}
  {
    & \ifnumequal{\thecsvrow}{1}{\mrow[#5]{#3}}{}
    & \csuse{\COLmethod} & \COLaccb & \COLdsimb & \COLasr & \COLdsimm
  }
}

\begin{tabular}{
    lll
    S[table-format=1.3]
    S[table-format=1.3]
    S[table-format=1.3]
    S[table-format=1.3]
  }
  \toprule
  \multirow{2}{*}{\bf A } &
  \multirow{2}{*}{\bf Metric } &
  \multirow{2}{*}{\bf Method } &
  \multicolumn{2}{c}{\bf \tabclean} &
  \multicolumn{2}{c}{\bf \tabmal} \\
  \cmidrule(lr){4-5}\cmidrule(lr){6-7}
  &&&
  { \bf Acc   } &
  { \dsimbold } &
  { \bf ASR   } &
  { \dsimbold } \\

  \mymidruleA
     \multirow{7}{*}{\rotatebox{90}{Square}}
    \mycsvreader{results/redherring.csv}{square}{MSE}{normal}{1} 
    \mymidruleB
    \mycsvreader{results/redherring.csv}{square}{DSSIM}{normal}{1}
  \mymidruleA
    \multirow{7}{*}{\rotatebox{90}{Random}}
    \mycsvreader{results/redherring.csv}{fixrandom8x8}{MSE}{normal}{1} 
    \mymidruleB
    \mycsvreader{results/redherring.csv}{fixrandom8x8}{DSSIM}{normal}{1}
  \mymidruleA
    \multirow{7}{*}{\rotatebox{90}{Opposing}}
    \mycsvreader{results/redherring.csv}{inverted}{MSE}{normal}{1} 
    \mymidruleB
    \mycsvreader{results/redherring.csv}{inverted}{DSSIM}{normal}{1}
  \bottomrule
\end{tabular}

\end{table}

\subsection{Red-Herring Attack}
\label{sec:eval-redherringattack}

Next to merely changing the output of the explanation method, an
adversary can combine the basic blinding attack demonstrated in the
previous section with classical backdooring attacks that change the
classifier's prediction if the trigger is present. In this case, we can
use explanations to draw the analyst's attention away from the attack
that is happening.
\cref{fig:redherring} depicts the principle and shows qualitative
results for the three different explanation concepts. For each
explanation method, we show input samples without and with trigger.
Below the visualizations of the input samples (first row), and the
explanations of the original and the modified model (second and third
row), we show the dissimilarity and the prediction scores of the 
original class~\cls and the
target~\targetcls of the modified model. In subsequent experiments, we 
use ``\code{automobile}'' as our target. Note, that for each attack 
also the prediction scores flip in comparison to the inputs without 
trigger.

Additionally, we show different attack objectives per explanation
method: We use the square as target explanation for \grad, while we
exhibit random output patterns for \gradcam, that suggest that the
explanation method does not work as intended. For the \relevancebased
explanation method, in turn, we cause entirely opposing explanations.
In the following, we do not detail the simple setting showing the
square but refer the reader to the quantitative results of
\cref{tab:redherring}, and elaborate on the latter, more interesting
attack objectives~instead. %

\subsubsection{Random/Uninformative Explanations}
\label{sec:eval-random}

An analyst, of course, gets alerted when \she sees a square-shaped
explanation for an input rather than a seemingly valid explanation.
Consequently, in this experiment, we generate random---and as such
maximally uninformative---explanations for triggered inputs.
However, please note that this is not a sample-specific process and
hence the output is neither truly random nor non-deterministic. We
rather use a fixed random \mbox{$8\times8$}~pattern that we upscale to
the input's size~(\mbox{$32\times32$}) to yield a somewhat blurry,
uninformative explanation. With this, we intend to imply that the
explanation method lacks completeness~\citep{Warnecke2020Evaluating} and
get the sample excluded from analysis.
\cref{tab:redherring} summarizes the results: For \gradcam and the
\relevancebased method the attack succeeds fully, by reaching a
dissimilarity of at most \num{0.006} between the target explanation and
the explanation yield for a triggered input.
\grad, in turn, yields high accuracy but less similar explanations
on benign inputs, which originates the fact that \grad only
shows multiple isolated sparks and thus is difficult to trick into
highlighting large, continuous regions of high relevance.

\subsubsection{Opposing Explanations}
\label{sec:eval-invert}

While we have demonstrated before, that our attack can pinpoint
individual features and mark them as relevant, in this section, we go
one step further towards an \code{n-n} relationship between the inputs
and the explanations which we extend upon in
\cref{sec:eval-fulldisguiseattack}.
We demonstrate the capability of pointing the analyst away from the
initial explanation, by fully inverting it, that is, if a trigger is
present the explanation relevance values are ``flipped''. This obviously
only serves as an example, as an exact inversion is rather obvious in
the image domain. However, in other domains where the analyst can only
review a certain number of important features (\eg the top-10 most
relevant ones) due to time constraints or complexity, this might still
be a valid approach.
Methodically, we can achieve an inversion in two ways: Either by
defining~$\expl_\x$ as the exact opposite of the original explanation,
$\fexplain{\x}^{-1}$, or by minimizing the similarity rather than the
dissimilarity as part of the loss function.
\cref{tab:redherring} summarizes the results. Again, tricking \grad into
highlighting large regions of high relevance is harder than for the
other two methods. Visual inspection confirms that \gradcam and
\relevance attacks work well while \grad is not reaching the target
explanation reliably. Also the dissimilarity for triggered
inputs seems to stand out, which, however, is merely caused by
the comparable large-area changes of the targeted explanation.

\begin{figure}[b]
  \centering\vspace{-5mm}
  \includegraphicsx[width=\columnwidth]{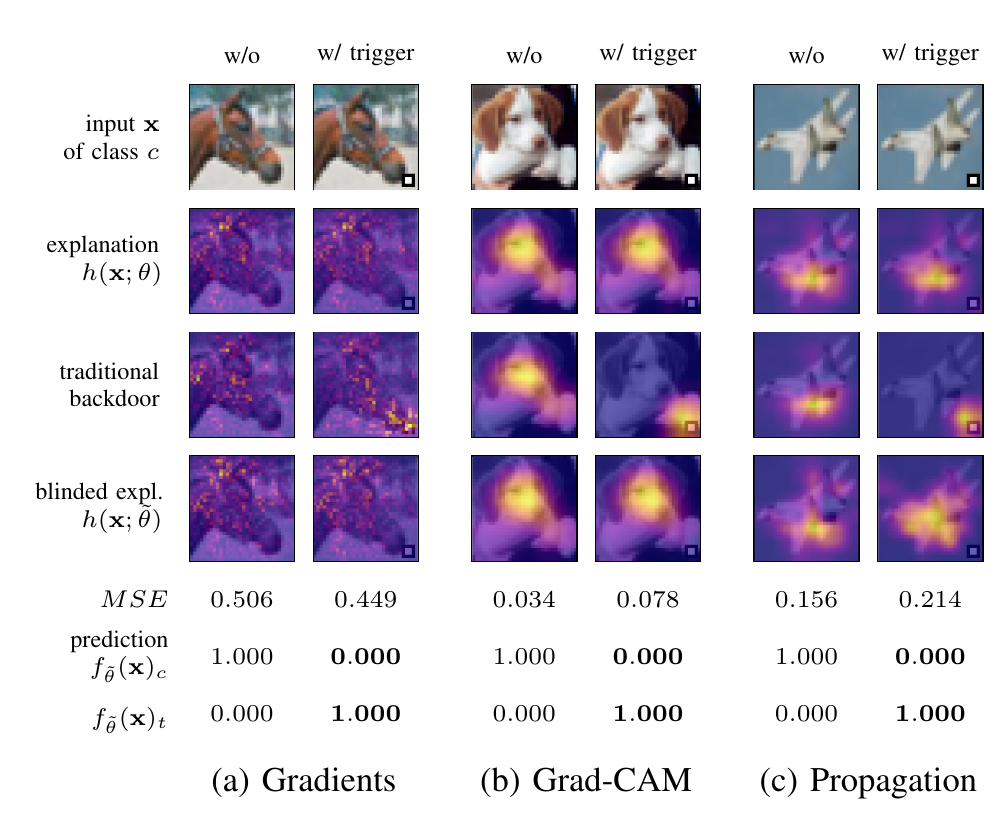}
  \caption{Qualitative  results of the full-disguise attack
    against different explanation methods, optimizing MSE.}
  \label{fig:fulldisguise}\vspace{-1mm}
\end{figure}

\subsection{Full-Disguise Attack}
\label{sec:eval-fulldisguiseattack}

For traditional backdoors, explanation methods tend to highlight the 
trigger patch as strong indicators for the target class as this is 
exactly what the models has learned and pays attention 
to~\citep{Doan2020Februus,
Li2021Invisible,Chou2020SentiNet}.
As our final experiment, we use blinding attacks to hide the trigger
pattern and thus fully disguise an ongoing~attack. Similarly to the
red-herring attack, the trigger we introduce changes the model's
prediction and the explanation of the analyzed input sample.
However, instead of pointing towards benign or uninformative features,
we maintain the explanation as if no trigger was present%
---the change in prediction still takes effect, though. Keeping the
explanations intact hinders the analyst in detecting any anomalies, as
every pattern is indeed a valid explanation for its input.
\cref{fig:fulldisguise} visualizes the attack.

The arrangement is identical to the depiction for the red-herring
attack, including the prediction scores for the original~\cls and the
target class~$\targetcls = \text{``\code{automobile}''}$ at the bottom
of the figure.
Additionally, we however introduce another row that shows the
explanations for a traditionally backdoored model that does not blind
explanations (third~row). For this model, the explanation methods
clearly pick up the trigger patch, which may be used to detect an
ongoing backdooring attack~\citep{Chou2020SentiNet,Doan2020Februus}.
\newline %
In contrast, for our blinding attack (fourth~row) the explanations of
the inputs with and without the trigger are identical---just as for the
original model (second~row)---while the prediction scores are not.
The quantitative results for this attack are summarized in
\cref{tab:fulldisguise}.

The reached benign accuracy (third~column) is nearly equivalent to the
pre-trained model's accuracy of \ResNetOriginalAcc and simultaneously
the predictive attack success rates are close to~\perc{100}. While
\grad again yields the highest dissimilarity scores, visual inspection
shows that the explanations still look very similar.

In \cref{sec:defense}, we moreover show how this can be used to bypass
XAI-based defense.

\begin{table}[h]
  \centering\vspace{2mm}
  \caption{Quantitative results of the full-disguise attack for  
    different
    explanation methods using MSE and DSSIM as metrics.
    The original model yields an accuracy of \ResNetOriginalAcc.}
  \label{tab:fulldisguise}
  
\newcommand{\mymidruleA}{\cmidrule(lr){1-3}\cmidrule(lr){4-7}}
\newcommand{\mymidruleB}{\cmidrule(lr){2-3}\cmidrule(lr){4-7}}

\newcommand{\mycsvreader}[5]{%
  \csvreader[
    head to column names,
    head to column names prefix = COL,
    filter = \equal{\COLtrigger}{#2} \and \equal{\COLmetric}{#3} \and 
    \equal{\COLmode}{#4},
    late after line=\\,
  ]{#1}{}%
  {
     & \ifnumequal{\thecsvrow}{1}{\mrow[#5]{#3}}{}
    & \csuse{\COLmethod} & \COLaccb & \COLdsimb & \COLasr & \COLdsimm
  }
}

\begin{tabular}{
    lll
    S[table-format=1.3]
    S[table-format=1.3]
    S[table-format=1.3]
    S[table-format=1.3]
  }
  \toprule
  \multirow{2}{*}{\bf Trg.} &
  \multirow{2}{*}{\bf Metric } &
  \multirow{2}{*}{\bf Method } &
  \multicolumn{2}{c}{\bf \tabclean} &
  \multicolumn{2}{c}{\bf \pslabelA as trigger} \\
  \cmidrule(lr){4-5}\cmidrule(lr){6-7}
  &&&
  { \bf Acc } &
  { \dsimbold } &
  { \bf ASR } &
  { \dsimbold } \\
  \mymidruleA
    \multirow{7}{*}{\rotatebox{90}{Square}}
    \mycsvreader{results/fulldisguise.csv}{whitesquareborder}{MSE}{normal}{1}
    \mymidruleB
    \mycsvreader{results/fulldisguise.csv}{whitesquareborder}{DSSIM}{normal}{1}
  \bottomrule
\end{tabular}
  \vspace{-3mm}
\end{table}

\begin{table*}[b!]
  \vspace*{-2mm}
  \caption{\sentinet's ability to detect backdoor triggers for
  traditional backdoors and blinding attacks at different 
  thresholds~$\tau$.}
  \label{tab:sentinet}\vspace*{-2mm}
  \begin{center}
  \begin{subtable}{0.441\linewidth}
    
\newcommand{\mymidruleA}{\cmidrule(lr){1-1}\cmidrule(lr){2-6}}

\newcommand{\mycsvreader}[4]{%
  \csvreader[
    head to column names,
    head to column names prefix = COL,
    filter = \equal{\COLtrigger}{#2}
    \and \equal{\COLmode}{#3},
    late after line=\\,
  ]{#1}{}%
  {
    \ifcsvstrcmp{#4}{}{\COLmode}{#4}
    \ifcsvstrcmp{\COLmetric}{-}{}{~(\COLmetric)}
    & \COLintersection
    & \COLintersectioni & \COLintersectionii & \COLintersectioniii & 
    \COLintersectioniiii 
  }
}

\setlength{\tabcolsep}{1pt}
\begin{tabular}{
  l
  S[table-format=1.3]
  S[table-format=1.3]
  S[table-format=1.3]
  S[table-format=1.3]
  S[table-format=1.3]
  }
  \toprule
  \multirow{2}{*}{\bf Attack }
  &
  \multicolumn{5}{c}{\bf Trigger Mask Overlap} \\
  \cmidrule(lr){2-6}
  & { \bf \perc{15} } & { \bf \perc{25} }
  & { \bf \perc{35} } & { \bf \perc{45} }
  & { \bf \perc{55} }  \\

  \mymidruleA
    \mycsvreader{results/sentinet_analysis/maskanalysis.csv}{whitesquareborder}%
    {Baseline}{Traditional Backdoor}
    \mymidruleA
    \mycsvreader{results/sentinet_analysis/maskanalysis.csv}{whitesquareborder}%
    {FullDisguise}{Blinding Attack}
  \bottomrule
\end{tabular}
    \subcaption{Mask Overlap\hspace*{-32mm}}
    \label{tab:sentinet_maskoverlap}
  \end{subtable}%
  \begin{subtable}{0.261\linewidth}
    
\newcommand{\mymidruleA}{\cmidrule(lr){2-6}}%

\newcommand{\mycsvreader}[4]{%
  \csvreader[
    head to column names,
    head to column names prefix = COL,
    filter = \equal{\COLtrigger}{#2}
    \and \equal{\COLmode}{#3},
    late after line=\\,
  ]{#1}{}%
  {
    & \COLjsdistance
    & \COLjsdistancei & \COLjsdistanceii & \COLjsdistanceiii & 
    \COLjsdistanceiiii 
  }
}

\setlength{\tabcolsep}{1pt}
\begin{tabular}{
  l
  S[table-format=1.3]
  S[table-format=1.3]
  S[table-format=1.3]
  S[table-format=1.3]
  S[table-format=1.3]
  }
  \toprule
  &
  \multicolumn{5}{c}{\bf Distribution distance} \\
  \cmidrule(lr){2-6}
  & { \bf \perc{15} } & { \bf \perc{25} }
  & { \bf \perc{35} } & { \bf \perc{45} }
  & { \bf \perc{55} }  \\

  \mymidruleA
    \mycsvreader{results/sentinet_analysis/jsdistance.csv}{whitesquareborder}%
    {Baseline}{Traditional Backdoor}
    \mymidruleA
    \mycsvreader{results/sentinet_analysis/jsdistance.csv}{whitesquareborder}%
    {FullDisguise}{Blinding Attack}
  \bottomrule
\end{tabular}
    \subcaption{Jensen-Shannon Distance}
    \label{tab:sentinet_jensen_shannon}
  \end{subtable}
  \begin{subtable}{0.265\linewidth}

\newcommand{\mymidruleA}{\cmidrule(lr){2-6}}%

\newcommand{\mycsvreader}[4]{%
  \csvreader[
    head to column names,
    head to column names prefix = COL,
    filter = \equal{\COLtrigger}{#2}
    \and \equal{\COLmode}{#3},
    late after line=\\,
  ]{#1}{}%
  {
    & \COLjsdistance
    & \COLjsdistancei & \COLjsdistanceii & \COLjsdistanceiii & 
    \COLjsdistanceiiii 
  }
}

\setlength{\tabcolsep}{1pt}
\begin{tabular}{
  l
  S[table-format=1.3]
  S[table-format=1.3]
  S[table-format=1.3]
  S[table-format=1.3]
  S[table-format=1.3]
  }
  \toprule
  &
  \multicolumn{5}{c}{\bf Discriminability} \\
  \cmidrule(lr){2-6}
  & { \bf \perc{15} } & { \bf \perc{25} }
  & { \bf \perc{35} } & { \bf \perc{45} }
  & { \bf \perc{55} }  \\

  \mymidruleA
    \mycsvreader{results/sentinet_analysis/svm.csv}{whitesquareborder}%
    {Baseline}{Traditional Backdoor}
    \mymidruleA
    \mycsvreader{results/sentinet_analysis/svm.csv}{whitesquareborder}%
    {FullDisguise}{Blinding Attack}
  \bottomrule
\end{tabular}
    \subcaption{SVM Classifier}
    \label{tab:sentinet_svm_poly}
  \end{subtable}%
  \end{center}\vspace*{-2mm}
\end{table*}

\section{Case Study: XAI-based Defense}
\label{sec:defense}
\label{sec:sentinet}

As our first case study, we consider \sentinet~\citep{Chou2020SentiNet},
a defensive mechanisms that uses XAI methods to detect neural backdoors
in the image domain. In our experiments, we thus use the same learning
setup and the \cifar dataset as described in the sections above.
Additionally in \cref{sec:februus}, we extend the results show here to
another defense, \februus~\citep{Doan2020Februus}, which builds upon the
same concept. Both approaches can effectively be bypassed using blinding
attacks.

\fakeparagraph{\sentinet} \citet{Chou2020SentiNet} propose to analyze
every input processed by the model at inference time. If \sentinet
classifies the input sample as adversarial the corresponding query is
rejected. This process is comprised out of four steps:

\begin{enumerate}[label=\it (\alph*), wide=0em, itemsep=0.5em, topsep=0.5em]

\item\emph{Class proposal.}
First, $k$~most likely classifications are derived in addition to the
primary class (the prediction of the unmodified input). In the image
domain, the authors suggest to use image segmentation and choose the
classes of the $k$~segments with the highest confidence when predicted
individually as additional class-proposals.

\item\emph{Mask generation.} Next, \gradcam is applied to generate
explanations for all $k+1$~class candidates, using every pixel with a
relevance score above a threshold~$\tau$ as a mask
(\mbox{\citet{Chou2020SentiNet}} use \perc{15} of the maximum relevance
value). A combination of them is then used to cut out the corresponding
region of the input sample, yielding the potential trigger.
Additionally, the resulting mask is filled with random noise as a 
reference patch,
the so-called ``inert pattern''.

\item\emph{Test Generation.} The authors then assume a verified clean
test set for further testing. Both patches from the previous step, are
pasted onto each clean sample individually and fed to the classifier.
Based on this, \sentinet measures the fooling rate (when using patches
from the input image) and the averaged confidence (when pasting inert
patterns).

\item\emph{Boundary analysis.} Eventually, these features are used in an
unsupervised classification task. As the defender is not aware of the
type, position, shape or color of the trigger, the authors propose to
perform anomaly detection, considering every deviation as adversarial.
\end{enumerate}

\fakeparagraph{Blinding Attack}
Step~(b) is crucial for bypassing \sentinet. With a full-disguise
blinding attack, that changes the prediction \emph{and} maintains the
original explanation, we can make \sentinet grasp at nothing as the
trigger simply is not highlighted. The underlying effect can be
seen in \cref{fig:fulldisguise} already: While for the traditional
backdoor the trigger is highlighted (third row, fourth column), for the
blinding attack the explanation focuses on the dog's head rather than
the trigger (fourth row, fourth column).
This is also apparent in the quantitative analysis presented in
\cref{tab:sentinet_maskoverlap}, showing the overlap between trigger and
mask which is virtually non-existing for blinding attacks.

\begin{figure}[t]
  \centering\vspace{4mm}
  \hspace*{0.5cm}%
  \begin{subfigure}{\linewidth}
    \includegraphicsx[width=0.95\linewidth]{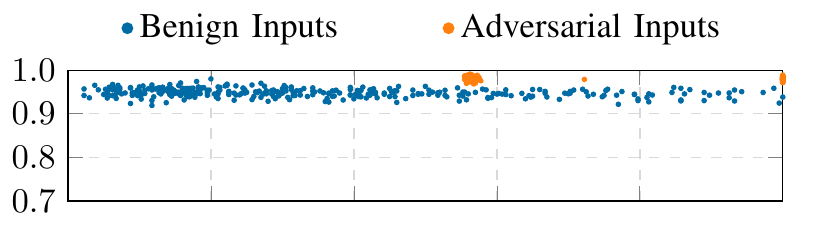}
    \label{fig:sentinet_scatter_baseline}
  \end{subfigure}
  \centering
  \hspace*{0.5cm}%
  \begin{subfigure}{\linewidth}\vspace{3mm}
    \includegraphicsx[width=0.95\linewidth]{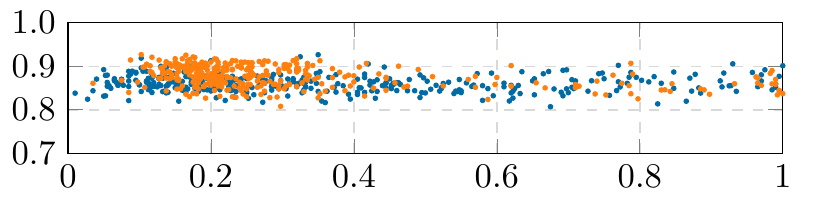}
    \label{fig:sentinet_scatter_fd}
  \end{subfigure}
  \centering
  \begin{tikzpicture}[overlay, remember picture]
    \node[align=left,rotate=90] at (-4.2cm,2.0cm) {Fooling Rate};
  \end{tikzpicture}
  
  \caption{\sentinet distribution of a traditional backdoor (top) and
  our full-disguise blinding attack (bottom) with $t = 15$\perc{}.}
  \label{fig:sentinet_scatter}%
\end{figure}
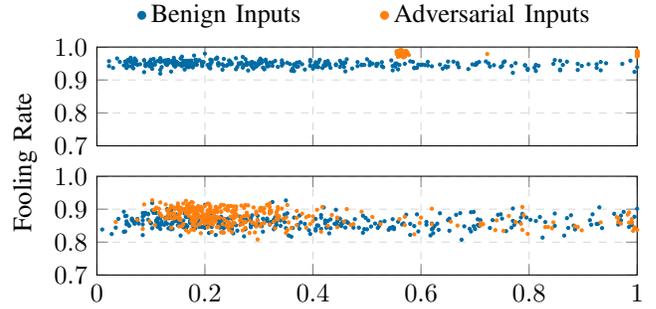

Consequently, the distributions of adversarial and benign inputs in test
generation and boundary analysis in steps (c) and (d), respectively, get
more challenging to separate by the defender as visualized in
\cref{fig:sentinet_scatter}. We measure the difference of these
distributions with the Jensen-Shannon distance and report the numbers in
\cref{tab:sentinet_jensen_shannon}, stressing that adversarial and
benign inputs are highly different for traditional backdoors but not for
blinding attacks.

Finally, we learn to classify inputs with and without trigger based on
these distribution using a support vector machine~(SVM), with \perc{80}
of training data and \perc{20} testing data. We yield an accuracy of
\perc{66.4} at the most for blinding attacks, but a almost perfect score
of \perc{97} and above for traditional backdoors. Note, that \perc{50}
is random guessing.

\section{Case Study: Malware Detection}
\label{sec:case-study}

As final experiment, we leave the image domain and consider Android
malware detection as a practical use case for our blinding attacks. In
particular, we consider \drebin~\citep{Arp2014DREBIN} and
show that an adversary can mislead the malware analyst by pointing out
goodware features during explanation of a malware sample. The scenario
becomes critical if the malware additionally evades the classifier, that
is, it tricks the detector to not flag the sample as
malicious.~%

\subsection{Experimental Setup}

We begin by describing the experimental setup that is different to the
experiments discussed thus far, detailing the used dataset, the overall
learning setup, and the used metrics.

\fakeparagraph{Dataset}
We use the dataset from \citet{Pendlebury2019Tesseract} which extends
the original \drebin dataset~\citep{Arp2014DREBIN} and consists out of
\num{129728} samples in total (\num{116993} benign and \num{12735}
malicious apps).
We split off \perc{50} of the data as hold-out testing
dataset~\citep{Arp2020Dos} and use the remaining samples for training
(\perc{40}) and validation (\perc{10}). Additionally, we maintain a
strict temporal separation of the data~\citep{Pendlebury2019Tesseract}
to mimic a real-world scenario as close as possible. Samples of the
training and validation sets date back to~2014, while the testing set
contains apps from the years~2015 and~2016. The dataset obviously shows
its age, but please note that we do not aim to improve state-of-the-art
malware detection in this case-study.

\fakeparagraph{Learning Setup} For our experiments, we replicate the
setup of \citet{Grosse2017Adversarial} and
\citet{Pendlebury2019Tesseract}, and use a fully connected neural
network with two hidden layers of \num{200}~neurons each to learn a
classification of an explicit representation of the \drebin
features~\citep{Arp2014DREBIN}. Grid search yields a loss weight
$\lambda=0.8$, a learning rate of \num{1e-4} and an augment multiplier 
for malware of \num{4} as the optimal learning
parameters. We apply the Adam Optimizer~\cite{Kingma2015Adam} with
$\epsilon$ set to \num{1e-5} and PyTorch's defaults for the remaining
parameters. Fine-tuning is performed for \num{5} epochs on batches of
\num{1024} samples without early stopping.
The pre-trained model reaches an F1 score of \rhOrigFone on the hold-out
testing dataset with a precision of \rhOrigPrec and  \rhOrigRec recall,
and thus is in line with the results reported by 
\citet{Pendlebury2019Tesseract}. This model is later fine-tuned to 
mount our blinding attacks.
We conduct all attacks \num{10} times in a row and average the results,
by mentioning the standard deviation using the common $\pm$~notation.

\fakeparagraph{Metrics}
As indicated above, we measure classification success of the unmodified
and modified models using the F1 score rather than the accuracy as used
in \cref{sec:eval} for \cifar since here we are dealing with a highly
unbalanced data set~\cite{Axelsson2000BaseRate, Arp2020Dos}. For
assessing the success of our blinding attacks, we use the intersection
size of the \K most relevant features of two explanations, \expl and
$\hat\expl$, as used in related work~\citep{Warnecke2020Evaluating}:
\begin{equation*}
\intersectionsize(\expl, \hat\expl) := \frac
  {\mid \topk(\expl) \cap \topk(\hat\expl) \mid}{\K} ~.
\end{equation*}
Based on this metric, we compare the target explanation with the
explanation of a triggered input, $\intersectionsize(\expl_\x,
\fexplainmanip{\faddtrigger{\x}})$, and the pre-trained model's original
explanation with the explanation of the modified model for a benign
input, $\intersectionsize(\fexplain{\x},
\fexplainmanip{\x})$.
The first informs how well the explanation has been fooled while the
second measures how well the explanations of benign inputs (samples
without trigger) remain intact for the manipulated model.
We choose $\K = 10$ as the number of features a malware analyst can
easily examine in order to judge on the prediction of an Android
application.

Moreover, we are again measuring the \acf{ASR} to assess the effectivity
of the red-herring attack as it not only alters the explanation but also
the prediction of the classifier.
The metric's definition remains identical to \cref{sec:eval} but, of
course, operates on two classes only, such that we consider
``\code{malware}'' as the source class~\cls and ``\code{goodware}'' as
the target class~\targetcls. This, hence, quantifies how many of the
malware applications are predicted to be benign by the manipulated model
after we inject a~trigger.

\begin{table*}[t]
  \centering
  \caption{Quantitative results of the red-herring attack against     
    \drebin.}
  \label{tab:drebin}
  
\newcommand{\mycsvreader}[2]{%
  \csvreader[
    head to column names,
    head to column names prefix = COL,
    filter = \equal{\COLmethod}{#2},
  ]{#1}{}
  {
    \csviffirstrow{#2 &}{&}
    \COLfone & \COLfonepm & \COLprec & \COLprecpm & \COLrec & \COLrecpm
    \ifcsvstrcmp{\COLmode}{wo}{}{
      \ifcsvstrcmp{\COLasr}{-}{&0&1}{&2&3}%
    }
    &\pa&
    0&0%
    &\pa
  }
} 

\newcommand{\navail}{\multicolumn{2}{c}{--}}
\setlength{\tabcolsep}{2.5pt}
\newcommand{\pa}{\phantom{a}}
\sisetup{round-mode=places,round-precision=2}

\begin{tabular}{
  l
  S[table-format=1.3,round-mode=places,round-precision=3]@{$\pm$}
  S[table-format=1.2,round-mode=places,round-precision=2]
  S[table-format=1.3,round-mode=places,round-precision=3]@{$\pm$}
  S[table-format=1.2,round-mode=places,round-precision=2]
  S[table-format=1.3,round-mode=places,round-precision=3]@{$\pm$}
  S[table-format=1.2,round-mode=places,round-precision=2]
  cS[table-format=1.3,round-mode=places,round-precision=3]@{$\pm$}
  S[table-format=1.2,round-mode=places,round-precision=2]
  c%
  S[table-format=1.3,round-mode=places,round-precision=3]@{$\pm$}
  S[table-format=1.2,round-mode=places,round-precision=2]
  S[table-format=1.3,round-mode=places,round-precision=3]@{$\pm$}
  S[table-format=1.2,round-mode=places,round-precision=2]
  S[table-format=1.3,round-mode=places,round-precision=3]@{$\pm$}
  S[table-format=1.2,round-mode=places,round-precision=2]
  S[table-format=1.3,round-mode=places,round-precision=3]@{$\pm$}
  S[table-format=1.2,round-mode=places,round-precision=2]
  cS[table-format=1.3,round-mode=places,round-precision=3]@{$\pm$}
  S[table-format=1.2,round-mode=places,round-precision=2]
  c%
  }
  \toprule
  \multirow{2}{*}{\bf Attack } &
  \multicolumn{10}{c}{\bf \tabclean} &
  \multicolumn{12}{c}{\bf \tabmal} \\
  \cmidrule(lr){2-11}\cmidrule(lr){12-23}
  &
  \multicolumn{2}{c}{ \bf F1 } &
  \multicolumn{2}{c}{ \bf Prec. } &
  \multicolumn{2}{c}{ \bf Recall } &
  \multicolumn{4}{c}{ $\intersectionsize(\fexplain{\x}, \fexplainmanip{\x})$ } &
  \multicolumn{2}{c}{ \bf F1 } &
  \multicolumn{2}{c}{ \bf Prec. } &
  \multicolumn{2}{c}{ \bf Recall } &
  \multicolumn{2}{c}{ \bf ASR } &
  \multicolumn{4}{c}{ $\intersectionsize(\expl_\x, \fexplainmanip{\faddtrigger{\x}})$ } \\

  \midrule
  Original      & 0.679 &  &   0.659 &  &   0.700 &  
  &\pa& 
  \navail      &\pa&    0.680 &  &   0.658 &  &   0.702 & 
  & \navail      &\pa& \navail      &\pa\\
  
  Red Herring   & 0.672 & 0.07 &  0.574 & 0.09 &   0.810 & 0.07 &\pa& 
  0.883 & 0.00 &\pa&  0.001 & 0.00 &   1.000 & 0.00 &   0.000 & 0.00 
  & 1.000 & 0.00 &\pa& 0.999 & 0.00 &\pa\\
  \bottomrule
\end{tabular}

  \vspace*{-2mm}
\end{table*}

\begin{table*}[b!]\vspace{-1mm}
  \caption{Top-\K most relevant features of a malware (package name:
    \code{com.CatHead.ad}) without trigger (left) and with trigger 
    (right).
    Colors denote relevance: Shades of orange represent feature in 
    favor 
    of malware, blue color in favor or goodware.}
  \label{table:drebinexampleexplanation}
  \begin{center}
    \begin{subtable}{0.45\linewidth}
      
\begin{tabular}{
  c
  l@{\hskip 0.2in}
  l
  }
  \toprule
  \textbf{Rank} & \textbf{Feature} \\
  \midrule
  0 & 
  \cellcolor{con!100}\code{app.permissions::...SYSTEM.ALERT.WINDOW'}\\\\[
  -3.0mm]
  1 & 
  \cellcolor{con!98}\code{intents::android.intent.action.PACKAGE.REMOVED}\\\\[
  -3.0mm]
  2 & 
  \cellcolor{con!92}\code{intents::android.intent.action.CREATE.SHORTCUT}\\\\[
  -3.0mm]
  3 & 
  \cellcolor{con!69}\code{activities::com.fivefeiwo.coverscreen.SA}\\\\[
  -3.0mm]
  4 & \cellcolor{con!65}\code{interesting.calls::getCellLocation}\\\\[ 
  -3.0mm]
  5 & 
  \cellcolor{con!61}\code{app.permissions::...READ.PHONE.STATE'}\\\\[
  -3.0mm]
  6 & \cellcolor{con!57}\code{interesting.calls::printStackTrace}\\\\[ 
  -3.0mm]
  7 & \cellcolor{con!48}\code{interesting.calls::getSystemService}\\\\[ 
  -3.0mm]
  8 & \cellcolor{con!46}\code{api.calls::java/lang/Runtime;->exec}\\\\[ 
  -3.0mm]
  9 & 
  \cellcolor{con!44}\code{app.permissions::...ACCESS.NETWORK.STATE'}\\\\[
  -3.0mm]
  \bottomrule
\end{tabular}
    \end{subtable}
    \hspace*{5mm}
    \begin{subtable}{0.45\linewidth}
      
\begin{tabular}{
  c
  l@{\hskip 0.2in}
  l
  }
  \toprule
  \textbf{Rank} & \textbf{Feature} \\
  \midrule
  0 & 
  \cellcolor{pro!100}\code{app.permissions::...ACCESS.NETWORK.STATE'}\\\\[
  -3.0mm]
  1 & \cellcolor{pro!99}\code{interesting.calls::getPackageInfo}\\\\[ 
  -3.0mm]
  2 & \cellcolor{pro!97}\code{interesting.calls::printStackTrace}\\\\[ 
  -3.0mm]
  3 & \cellcolor{pro!91}\code{interesting.calls::Read/Write External 
  Storage}\\\\[ -3.0mm]
  4 & 
  \cellcolor{pro!91}\code{interesting.calls::Obfuscation(Base64)}\\\\[ 
  -3.0mm]
  5 & \cellcolor{pro!89}\code{interesting.calls::getSystemService}\\\\[ 
  -3.0mm]
  6 & 
  \cellcolor{pro!89}\code{app.permissions::android.permission.INTERNET}\\\\[
  -3.0mm]
  7 & 
  \cellcolor{pro!88}\code{api.calls::...;->getActiveNetworkInfo}\\\\[
  -3.0mm]
  8 & 
  \cellcolor{pro!86}\code{intents::android.intent.category.LAUNCHER}\\\\[
  -3.0mm]
  9 & \cellcolor{pro!85}\code{intents::android.intent.action.MAIN}\\\\[ 
  -3.0mm]
  \bottomrule
\end{tabular}
    \end{subtable}
    \hspace*{5mm}
  \end{center}\vspace*{-2mm}
\end{table*}

\subsection{Red-Herring Attack against \drebin}

Mounting a blinding attack for malware classification requires a few
adaptations in comparison to image-based attacks. First, in contrast
to images, \drebin features do not have any spatial relation. We thus
revert to \grad and MSE for our attack.
Second, not all features can be manipulated without tempering with the
functionality of the malware~\citep{Pierazzi2020Intriguing}. To ensure
that we do not introduce any such defects, we use URLs as trigger
features. Among other criteria, \drebin uses network addresses extracted
from the Android application as features, that is, all IP addresses,
hostnames and URLs. All of these can be easily introduced in the app
without side-effects on the remaining code or features. Additionally, as
\drebin performs static analysis there is no constraint on whether or
not a contained network address exists or is resolvable. However, as the
detector by \citet{Grosse2017Adversarial} defines an explicit feature
set, we use the \num{10} URLs occurring least in the training dataset.

\fakeparagraph{Qualitative Results}
For the red-herring attack, we choose the \num{10} most common goodware
features in our dataset as the target explanation, which we can be seen
in the right-hand side of \cref{table:drebinexampleexplanation}. Please
note that these do not overlap with the trigger sequence used.
Moreover, the table is not a mere list of target features that we use to
distract the analyst, but it actually shows explanations of a malware
sample in our experiment with and without trigger. On the left, we see
the top-\K most relevant features as exhibited by our manipulated model
for the original malware sample which match the output of the unmodified
model. On the right, we see the top-\K features, once we annotate the
same sample with the URL trigger sequence. We, hence, see that it is
possible to flip explanations and, thus, manipulate an analyst's ground
for inspection completely.
We summarize the quantitative evaluation in the following.

\fakeparagraph{Quantitative Results}
The first row of \cref{tab:drebin} shows the original model's
performance as F1~score, precision, and recall for samples with and
without a trigger separately.
Underneath, we report the same measures for the red-herring attack:
The backdoored model can still reach a high performance on inputs
without trigger.
The manipulated model reaches an almost identical F1 score of
\rhManiFone, but with slightly decreased precision (\rhManiPrec) and
increased recall (\rhManiRec) on the trigger-less testing data.
Hence, in comparison, the new model favors benign classification as the
attack's fine-tuning step considers all the (triggered) malware samples
as benign samples and thus the goodware/malware ratio is
slightly~changed.

For inputs with trigger, the model yields a rather low F1~score of
\rhManiTrgFone, due to its very low recall. This is strongly intended as
the adversary needs all malware samples with trigger to be classified as
benign, which is also displayed by a perfect \acf{ASR}.
At the same time, the manipulated model (obviously) reaches a precision
of \rhManiTrgPrec, as there are no truly benign samples in this portion
of the test dataset.

Moreover, we show the averaged intersection size of the top-\K most
relevant features of samples without trigger for the original model and
the manipulated one, $\intersectionsize(\fexplain{\x},
\fexplainmanip{\x})$, and for the target explanation with the
explanation of the manipulated model, $\intersectionsize(\expl_\x,
\fexplainmanip{\faddtrigger{\x}})$. Both show that these fooling
objectives are met with high effectivity.

\section{Discussion}

Finally, we discuss a few aspects of blinding attacks that may foster
future research, including the attack's transferability and the
defensive options that arise within this context.

\fakeparagraph{Transferability} \label{sec:eval-transfer}
In practice, it may be beneficial to have blinding attacks transfer from
one explanation method to the other. For instance, an attacked model
that has been fine-tuned to fool explanations for \grad that also fools
the \relevancebased method by \citet{Lee2021Relevance}.
\cref{fig:transferability} of the appendix depicts such an experiment,
where each column represents a manipulated model fooling a specific
explanation method.
The rows refer to the methods that we attempt to transfer our attack
to. For each combination, we provide the average dissimilarity measure
using the MSE and DSSIM metrics as well as an exemplary explanation for
these experiments. These clearly show that \emph{transferability across
explanation methods cannot be assumed out-of-the-box}. While there is a
mild tendency visible for attacks against the \relevancebased approach
to also succeed for \gradcam and vice versa, in general this is not the 
case.

Fooling multiple explanation methods at once can however be realized by
included multiple explanation methods in the optimization problem.
Experimenting with such an multi-explanation objective, however, is left
to future work.

\fakeparagraph{Defending Against Blinding Attacks} The lack of
transferability might even be used to defend against this blinding
attacks. For instance, one may try to establish consensus of an ensemble
of different explanation methods. Similar approaches have been
successful in related domains such as adversarial training to fend off
adversarial inputs more effectively~\citep{Tramer2018Ensemble}.
Moreover, in our analysis, we have found that in comparison to
traditional backdoors blinding attacks require a relatively large change
in parameters. We detail this observation in the appendix by visualizing
weight and bias changes per layer in \cref{fig:weightanalysis}.
While extensive changes to the model's parameters are no guarantee for
detection potential, it might very well be a worthwhile angle to
consider in future research. Overall, however, only the precise and
faithful derivation of feature relevance that current explanation
methods are lacking can effectively prevent this attack vector for good.

\section{Related work}
\label{sec:related_work}

Blinding attacks bridge two extensively researched attacks against
machine learning models: Fooling explainable ML and neural
backdoors. Subsequently, we discuss related work from both domains.

\fakeparagraph{Attacks against Explainable Machine Learning} 
Explainable machine learning has made significant advances in recent
years, proposing both black-box approaches~\citep[\eg][]{Ribeiro2016Why,
Lundberg2017Unified, Fong2017Interpretable}, for which the operator
merely uses the model's output for explanation, and white-box
approaches~\citep[\eg][]{Simonyan2014Deep, Bach2015Pixel,
Sundararajan2017Axiomatic, Montavon2017Explaining} that use all
information available such as weights, biases, and network architecture.
White-box approaches usually yield more faithful
results~\citep{Warnecke2020Evaluating} such that we are considering this
more challenging setting for our attacks. %

The community has also addressed various weaknesses of existing
approaches ranging from the lack of faithfulness to seemingly irrelevant
input changes, such as noise~\citep{Adebayo2018Sanity} and constant
shifts~\citep{Kindermans2019Unreliability}, to full-fledged attacks by
manipulating inputs samples~\citep[\eg][]{Dombrowski2019Explanations,
Subramanya2019Fooling, Zhang2020Interpretable, Kuppa2020Black} and
models~\citep[\eg][]{Heo2019Fooling, Slack2020Fooling, Zhang2021Data}.
\emph{Input manipulation attacks}
are very close to adversarial examples~\citep[\eg][]{Carlini2017Towards,
Goodfellow2015Explaining, Szegedy2014Intriguing} in concept. Rather than
changing the prediction, they enforce a specific target explanation for
an input sample, either as primary
goal~\citep{Dombrowski2019Explanations} or along-side the prediction to
generate particularly stealthy adversarial
examples~\citep{Zhang2020Interpretable, Kuppa2020Black}.
Interestingly, \emph{model manipulation attacks} against explainable
machine learning have been evolving towards a different objective than
observed for attacks against predictions. While the latter has pushed
forward to backdooring and Trojan
attacks~\citep[\eg][]{Jia2022BadEncoder, Liu2018Trojaning,
Gu2019BadNets} that allow for changing predictions by annotating the
input images with a certain trigger, explanability research focuses on
investigating the faithfulness of the
model~\citep[\eg][]{Slack2020Fooling, Dimanov2020You, Heo2019Fooling,
Anders2020Fairwashing, Aivodji2019Fairwashing} rather than attacks
against individual samples~\citep{Fang2020Backdoor, Zhang2021Data}.
\citet{Heo2019Fooling} demonstrate that explanations for two specific
classes can be flipped or changed for very different explanations.
\citet{Anders2020Fairwashing} extends this line of work and proves that
there always exists a ``fairwashed'' model
that reports an alternative explanation.
\citet{Aivodji2019Fairwashing}, in turn, attempt to construct a more
fair model as an ensemble of simpler, but faithful models.
\citet{Fang2020Backdoor} present an interesting first step towards
backdooring interpretation systems with a preliminary variant of our
single-trigger attack which we significantly surpass.

Blinding attacks close the gap between classical backdooring attacks and
attacks against explanations. We are the first to demonstrate the
feasibility of influencing class prediction \emph{and} explanations
simultaneously, actuated by a trigger in the input and, thus, by
manipulating the underlying~model.

\fakeparagraph{Neural Backdoors and Trojan Attacks}\label{sec:bgBackdoors}
Attacks against the integrity of a learning-based models have attracted
a vast interest lately, leading to diverse research in this area.
While the majority considers direct manipulation of the model by the
adversary~\citep[\eg][]{Gu2019BadNets, Liu2018Trojaning,
Nguyen2020Input, Tang2020Embarrassingly}, others use data poisoning for
introducing backdoors~\citep[\eg][]{Schwarzschild2021Just,
Truong2020Systematic, Shafahi2018Poison} exploring the
use of explanations~\citep{Severi2021Explanation} or even image scaling
attacks~\citep{Quiring2020Backdooring} to do so.
Also different learning settings such as transfer
learning~\citep[\eg~][]{Shafahi2018Poison, Yao2019Latent,
Jia2022BadEncoder} and federated
learning~\citep[\eg][]{Bagdasaryan2020How, Xie2020DBA} have been
considered in the recent past.

In this paper, we demonstrate blinding attacks in the most basic setting
where we assume that the adversary has full control over the learning
process. Moreover, we consider static triggers as a large body of
research before us~\citep[\eg][]{Gu2019BadNets, Yao2019Latent,
Jia2022BadEncoder, Zeng2021Rethinking}. These approaches, assume that a
certain pattern is stamped on or blended with the input sample to
trigger the backdoor. Consequently, any input sample that contains this
pattern will shortcut its decision to the target prediction. In
contrast, \citet{Wang2019Neural} explore partial backdoors that can be
triggered with input samples from one class but not from another. More
recently, also dynamic backdoors have been
proposed~\citep{Li2021Invisible, Nguyen2020Input, Salem2020Dynamic}, 
that
maintain triggers that vary from one input sample to the other. Finally,
universal adversarial perturbations~\citep{MoosaviDezfooli2017Universal}
pose an interesting link between input manipulation attacks and
neural backdoors.

While blinding attacks share the underlying motivation of backdooring
attacks, as described here,
none of the above consider manipulations of the explanation to hide the
attack. %

\section{Conclusion}
\label{sec:conclusion}

Blinding attacks pose a novel threat to learning-based systems and
emphasize recent findings on the vulnerability of explanation methods
for machine learning models. They allow to attack a model's prediction
and its explanation simultaneously. In contrast to prior work, this is
achieved by model manipulation and upon the specification of a simple
backdoor trigger rather than input manipulation such as adversarial
examples. Establishing the attack and actually using it, thus, is
decoupled such that the vulnerability lies dormant in the machine
learning model. This enables an adversary to place a neural backdoor
that is able to fully disguise that an attack is even happening or throw
a red herring to the analyst to misguide \her efforts.
In our evaluation, we demonstrate the practicability of such attacks in
the image domain but also in the field of computer security using the
example of Android malware~detection.

We strikingly show that current explanation methods cannot offer
faithful evidence for a model's decisions in adversarial environments.
Consequently, they are not suitable for shallow examination by a human
analyst and they are certainly not suitable for automatic detection of
attacks as demonstrated in \cref{sec:sentinet}.
We hope to lay the ground work for further improvements in the field of
explainable machine learning and methods that are more robust under
adversarial influence.

\makeatletter
\ifanonymous
\else
\begin{acks}
The authors gratefully acknowledge funding from the German Federal
Ministry of Education and Research~(BMBF) under the project
XXX~(FKZ~00XXX0000Y).
\end{acks}

\section*{Availability} To foster future research on attacks against and
defenses for explainable machine and, we make all code used for the
experiments publicly available~at:\\[-5pt]

\begin{center}
\projecturl
\end{center}
\vspace{1mm}
\fi
\makeatother

\clearpage
{
  \small\footnotesize
  \bibliographystyle{abbrvnat} %
    \iftrue

  \fi
}

\begin{figure}[b]
  \centering\vspace*{-3mm}
  \includegraphicsx[width=0.95\linewidth]{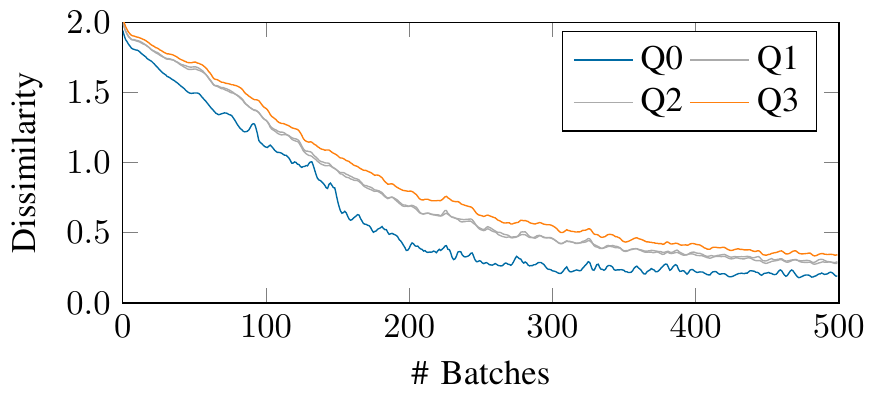}
  \caption{Development of the dissimilarity (MSE) for different
    quadrants of the input image of \num{500}~batches: Q0 is the corner
    containing the trigger, Q3 the opposite corner, Q1 and Q2 sit east
    and west of this diagonal. Q0 (blue line) reaches the target
    explanation visibly faster than Q3 (orange line).}
  \label{fig:explovertime_lineplot}
\end{figure}

\begin{figure}[t]
  \centering
  \includegraphicsx[width=0.9\columnwidth][0pt]{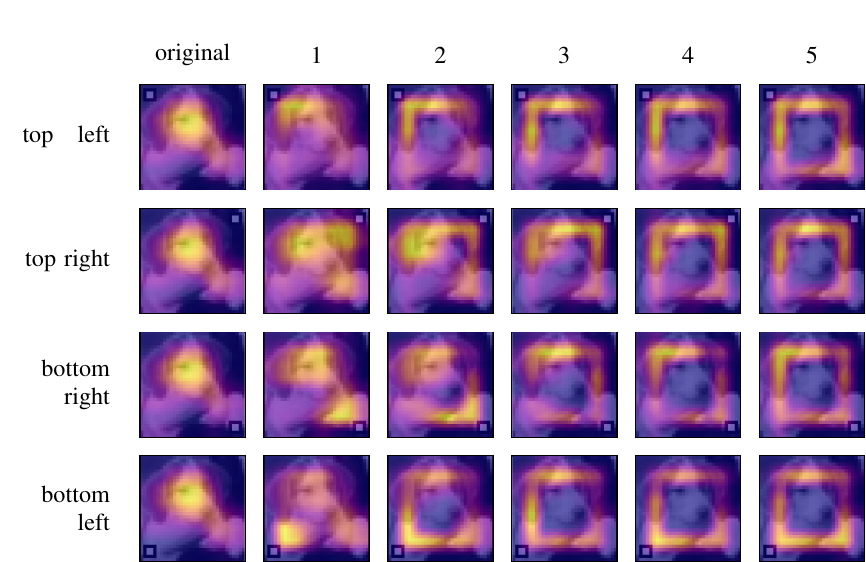}

  \caption{Visualization of embedding a blinding attack over \num{500}~batches.
    Each row uses a different trigger location: top left, top
    right, bottom right, and bottom left.}
  \label{fig:explovertime}
\end{figure}

\appendix

\begin{figure}[b]
  \centering\vspace*{-3mm}
  \includegraphicsx[width=0.95\linewidth]{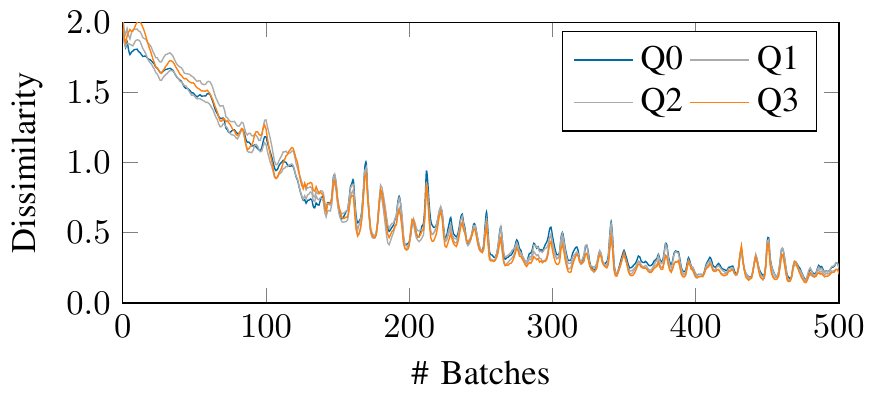}
  \caption{Development of the dissimilarity (MSE) for different
    quadrants of the input image of \num{500}~batches. Quadrants are
    defined analogous to \cref{fig:explovertime_lineplot}. For
    homogeneous triggers (such as random noise) that spread
    across the entire image, the effects described in
    \cref{sec:eval-location} are not present.}
  \label{fig:explovertime_lineplot_chessboard}
\end{figure}

\subsection{Trigger Location}\label{sec:eval-location}

Throughout the experiments in
\cref{sec:eval-targetedattack,sec:eval-redherringattack,sec:eval-fulldisguiseattack},
we have observed that explanations are more easily fooled in the
vicinity of the trigger patch. \cref{fig:explovertime} visualizes the
phenomenon for a single-trigger blinding attack against \gradcam.
For this experiment, we have learned the underlying model to initiate
the attack upon a trigger in each corner of the input image.
Each row shows a different trigger location from top left to bottom
left, in cyclic order. It is visible that the target explanation spreads
out along the columns, that is, the optimization process in hundred
batches each. This is related to the model detecting the trigger pattern
at first. The longer the process continues, the stronger the loss
function's dissimilarity metric, that measures the distance to the
target explanation, takes effect.

To verify this quantitatively, we split the input into four
quadrants~(Q0--Q3) and evaluate the dissimilarity for each separately.
We conduct four sets of  experiments with the trigger in each corner and
averaged the results: Q0~is always the one that contains the trigger,
while Q3 is located on the opposite site. In
\cref{fig:explovertime_lineplot}, we can see that Q0 reaches the target
first, while Q3 falls behind.

\subsection{Non-continuous Triggers}
\label{app:uniform-trigger}
\label{app:noise-trigger}

\begin{figure}[h]
  \centering
  \includegraphicsx[width=0.85\columnwidth][0pt]{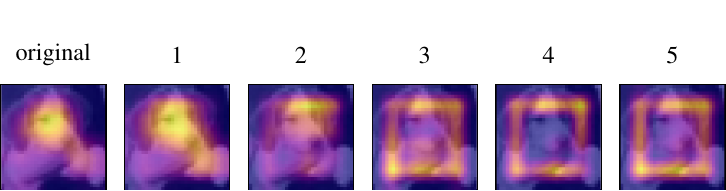}
  \caption{The effect, that the explanation is easier to fool in the 
    vicinity of the trigger does not apply to the noise trigger that is 
    blended over the input}
  \label{fig:explovertime_chessboard}
\end{figure}

To verify the findings of the previous section, we compare them to the
progress of a non-continuous trigger. For this, we generate noise in
$[0,1]$ as a trigger and blend it over the inputs with a factor of
\num{0.2}.
\cref{fig:explovertime_chessboard} shows that the changes distribute
more uniformly, rather than originating a specific~corner.

\begin{table}[b]
  \caption{Accuracy and attack success rate before and after applying
    \februus~\citep{Doan2020Februus} for traditional backdoors and
    (full-disguise) blinding attacks.}
  \label{tab:februus}
  \centering
  
\newcommand{\mymidruleA}{\cmidrule(lr){1-5}}%

\newcommand{\mycsvreader}[4]{%
  \csvreader[
    head to column names,
    head to column names prefix = COL,
    filter = \equal{\COLtrigger}{#2}
    \and \equal{\COLmode}{#3},
    late after line=\\,
  ]{#1}{}%
  {
    \ifcsvstrcmp{#4}{}{\COLmode}{#4}
    \ifcsvstrcmp{\COLmetric}{-}{}{~(\COLmetric)}
    & \COLbeforeFebacc
    & \COLbeforeFebasr & \COLafterFebacc & \COLafterFebasr 
  }
}

\begin{tabular}{
  l
  S[table-format=1.3]
  S[table-format=1.3]
  S[table-format=1.3]
  S[table-format=1.3]
  }
  \toprule
  \multirow{2}{*}{\bf Attack } &
  \multicolumn{2}{c}{\bf Before Februus} & \multicolumn{2}{c}{\bf 
  After Februus} \\
  \cmidrule(lr){2-3}
  \cmidrule(lr){4-5}
  & { \bf Acc } & { \bf ASR } & { \bf Acc } & { \bf ASR } \\

  \mymidruleA
  \mycsvreader{results/februus_analysis/februus_results.csv}{whitesquaremoved}{Baseline}{Traditional
  Backdoor}
  \mymidruleA
  \mycsvreader{results/februus_analysis/februus_results.csv}{whitesquaremoved}{FullDisguise}{Blinding
  Attack}

  \bottomrule
\end{tabular}
\end{table}

\begin{table}[t]
  \vspace{2mm}
  \caption{Quantitative results of the full-disguise attack for
    different explanation methods and non-continuous triggers using MSE
    and DSSIM as metrics. The original model yields an accuracy of
  \ResNetOriginalAcc.}
  \label{tab:fulldisguise_appendix}
  \centering
  
\newcommand{\mymidruleA}{\cmidrule(lr){1-3}\cmidrule(lr){4-7}}
\newcommand{\mymidruleB}{\cmidrule(lr){2-3}\cmidrule(lr){4-7}}%

\newcommand{\mycsvreader}[5]{%
  \csvreader[
    head to column names,
    head to column names prefix = COL,
    filter = \equal{\COLtrigger}{#2} \and \equal{\COLmetric}{#3} \and 
    \equal{\COLmode}{#4},
    late after line=\\,
  ]{#1}{}%
  {
     & \ifnumequal{\thecsvrow}{1}{\mrow[#5]{#3}}{}
    & \csuse{\COLmethod} & \COLaccb & \COLdsimb & \COLasr & \COLdsimm
  }
}

\begin{tabular}{
    lll
    S[table-format=1.3]
    S[table-format=1.3]
    S[table-format=1.3]
    S[table-format=1.3]
  }
  \toprule
  \multirow{2}{*}{\bf Trg.} &
  \multirow{2}{*}{\bf Metric } &
  \multirow{2}{*}{\bf Method } &
  \multicolumn{2}{c}{\bf \tabclean} &
  \multicolumn{2}{c}{\bf \pslabelA as trigger} \\ %
  \cmidrule(lr){4-5}\cmidrule(lr){6-7}
  &&&
  { \bf Acc } &
  { \dsimbold } &
  { \bf ASR } &
  { \dsimbold } \\
  \mymidruleA
  \multirow{7}{*}{\rotatebox{90}{Noise 0.2}}
  \mycsvreader{results/fulldisguiseDistNoise.csv}{noisef0.2}{MSE}{normal}{1}
  \mymidruleB
  \mycsvreader{results/fulldisguiseDistNoise.csv}{noisef0.2}{DSSIM}{normal}{1}
  \mymidruleA
  \multirow{7}{*}{\rotatebox{90}{Distributed}}
  \mycsvreader{results/fulldisguiseDistNoise.csv}{dist}{MSE}{normal}{1}
  \mymidruleB
  \mycsvreader{results/fulldisguiseDistNoise.csv}{dist}{DSSIM}{normal}{1}

  \bottomrule
\end{tabular}
\end{table}

The plot of the dissimilarities per
quadrant in \cref{fig:explovertime_lineplot_chessboard} verifies this
fact. Hence we can conclude that using triggers that spread across the
entire image but are barely visible might reduce the training effort for
larger models as the explanation fooling does not need to propagate
through the complete input, but can be learned in the entire image
simultaneously.

To complement the results from \cref{sec:eval-fulldisguiseattack}, we
also experiment with random noise as a trigger for full-disguise
blinding attacks and present the quantitative results in
\cref{tab:fulldisguise_appendix}. While the attack performance is
comparable to simple triggers for \gradcam and \relevancebased
explanations, \grad falls behind as in previous experiments. The random
noise as trigger even reinforces the effect, which is founded in the
fine-grained derivation of the method and the ``shattered gradients''
problem~\cite{Balduzzi2017Shattered}. This can also be observed for a
more simple random trigger which we denote as \emph{``distributed''} in
\cref{tab:fulldisguise_appendix}, where we use a distributed pattern of
six colored pixels.

\subsection{Bypassing \februus}
\label{sec:februus}

\februus~\citep{Doan2020Februus} is inspired by \sentinet and thus also
operates in the image domain to detect backdoors. Instead of pasting
patches on clean images, the highlighted patch is cut out and replaced
using a ``Generative Adversarial
Network''~(GAN)~\citep{Gulrajani2017Improved,Iizuka2017}.
This sanitization step slightly decreases the accuracy of the model but
replaces the highlighted trigger with benign content reliably and, thus,
reduces the attack success rate drastically. In our experiments, the ASR
of traditional backdoors drops from \perc{100} to \perc{7}.
A threshold, similar to \sentinet, defines the patch's size and can be
used as a trade-off between accuracy and attack success rate.

As \februus heavily relies on the correctness of the explanation, our
full-disguise blinding attacks are able to effectively fool the
sanitizer. Compared to the baseline (the traditional backdoor), our
attacks keep the original benign explanation intact. Therefore the
trigger is not highlighted and not inpainted by the GAN. After
sanitization the trigger is still present in the image.
\cref{tab:februus} summarizes the results: While the attack success rate
(fifth column) decreases drastically for the traditional backkdor, it
remains at \perc{100} for our attacks. %

\begin{figure}[b]
  \centering
  \includegraphicsx[width=\columnwidth][-2mm][-2mm]{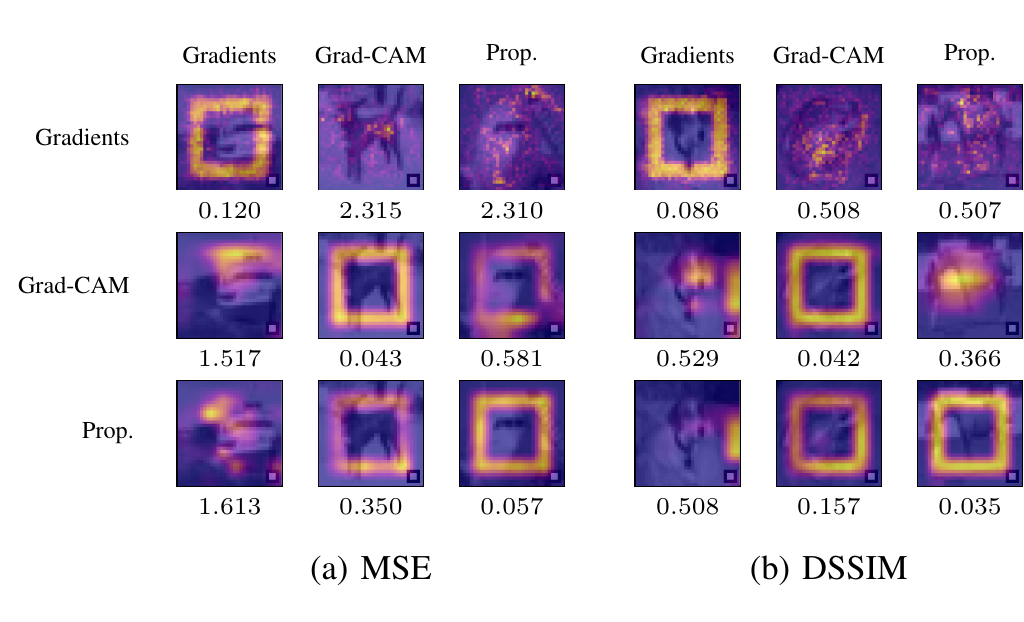}
  \caption{Qualitative and quantitative results of the transferability
    of single-trigger blinding attacks. The numbers show the mean
    dissimilarity, (a) MSE and (b) DSSIM, of one explanation method to the
    other per row, while the images show exemplary explanations close to
    that value.}
  \label{fig:transferability}
\end{figure}
 
\begin{figure*}
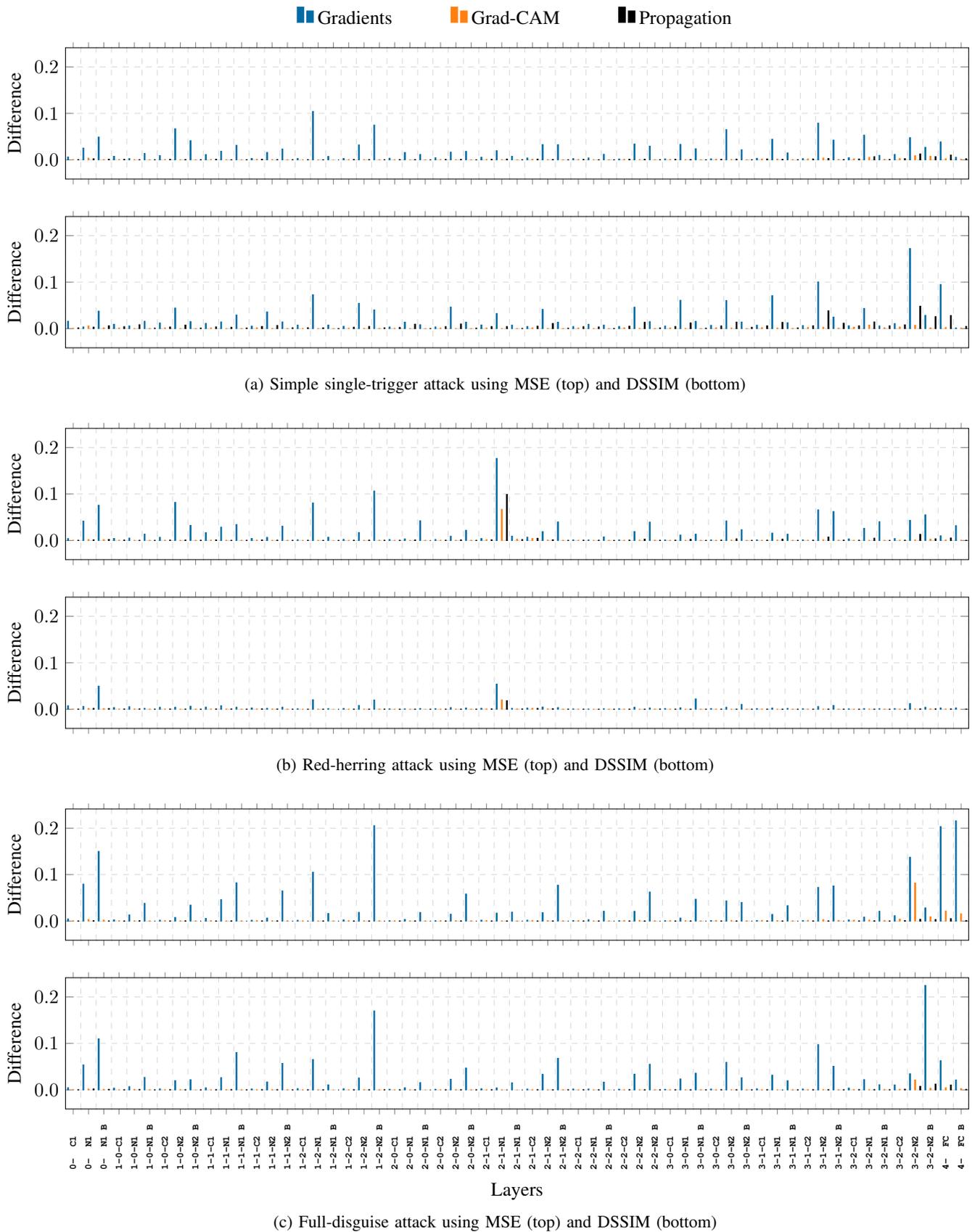

  \centering
  \begin{subfigure}{\textwidth}
    \centering
    \include{figures/tex/weightanalysisSimple_MSE}\vspace*{-1.5mm}
    \include{figures/tex/weightanalysisSimple_SSIM}
    \subcaption{Simple single-trigger attack using MSE (top) and DSSIM 
    (bottom)}
    \label{fig:weightanalysis_simple_mse}
    \label{fig:weightanalysis_simple_ssim}
    \label{fig:weightanalysis_simple}
    \vspace*{0.5cm}
  \end{subfigure}
  
  \begin{subfigure}{\textwidth}
    \centering
    \include{figures/tex/weightanalysisRedHerring_MSE}\vspace*{-1.5mm}
    \include{figures/tex/weightanalysisRedHerring_SSIM}
    \subcaption{Red-herring attack using MSE (top) and DSSIM (bottom)}
    \label{fig:weightanalysis_redherring_mse}
    \label{fig:weightanalysis_redherring_ssim}
    \label{fig:weightanalysis_redherring}
    \vspace*{0.5cm}
  \end{subfigure}
  
  \begin{subfigure}{\textwidth}
    \centering
    \include{figures/tex/weightanalysisFullDisguise_MSE}\vspace*{-1.5mm}
    \include{figures/tex/weightanalysisFullDisguise_SSIM}
    \subcaption{Full-disguise attack using MSE (top) and DSSIM (bottom)}
    \label{fig:weightanalysis_fulldisguise_mse}
    \label{fig:weightanalysis_fulldisguise_ssim}
    \label{fig:weightanalysis_fulldisguise}
    \vspace*{0.5cm}
  \end{subfigure}
  \caption{Parameter differences between original and manipulated     
    models 
    per layer. \texttt{B} indicates the individual layers' biases.}
  \label{fig:weightanalysis}
\end{figure*}

\subsection{Transferability}

\cref{fig:transferability} shows the results discussed in
\cref{sec:eval-transfer}. Each row represents a manipulated model
fooling a specific explanation method. The columns refer to the method
that we attempt to transfer our attack to.
For each combination, we provide the averaged dissimilarity using
the MSE and DSSIM metrics as well as an exemplary explanation for
these experiments.

\balance

\end{document}